\documentclass{article}

\usepackage{arxiv}
\usepackage[utf8]{inputenc} % allow utf-8 input
\usepackage[T1]{fontenc}    % use 8-bit T1 fonts
\usepackage{hyperref}       % hyperlinks
\usepackage{url}            % simple URL typesetting
\usepackage{booktabs}       % professional-quality tables
\usepackage{amsfonts}       % blackboard math symbols
\usepackage[numbers]{natbib}
\usepackage{amsmath, enumitem}
\usepackage{amssymb}
\usepackage{upgreek}
\usepackage{lineno}
\usepackage{soul}
\usepackage{fixltx2e}

\usepackage{nccmath}
\usepackage{ifthen}
\usepackage{color}
\usepackage{graphicx}
\usepackage{algorithm}
\usepackage{algorithmic}
\usepackage{setspace}
\usepackage{multirow}
\usepackage[T1]{fontenc}
\graphicspath{{Fig/}}

\title{Linearised versus Nonlinear Estimates of Uncertainty in Full Waveform Inversion}
\date{} 					% Or removing it

\author{
	Xuebin Zhao \\
	School of Geosciences \\
	University of Edinburgh\\
	Edinburgh, Unite Kingdom \\
	\And
	Andrew Curtis \\
	School of Geosciences \\
	University of Edinburgh\\
	Edinburgh, Unite Kingdom \\
}

\begin{document}
	\maketitle

\begin{abstract}
Seismic full waveform inversion (FWI) is a powerful technique that uses seismic waveform data to generate high resolution images of the Earth's interior. However, significant uncertainty exists in all FWI solutions due to imperfect acquisition geometries, inherent noise in the data, nonlinearity of the forward problem, and the under-determined nature of real-world tomographic problems in which the target is heterogeneous over all length scales. Probabilistic Bayesian FWI addresses this non-uniqueness by estimating the entire family of possible model solutions and thus the solution uncertainty, described by the so-called \textit{posterior} probability density function (pdf) over model parameter values. The posterior pdf can be estimated using nonlinear inversion methods to quantify full uncertainties, including those created by the nonlinearity in the physics. Alternatively, by linearising the physics relating parameters and observations around a chosen reference model solution, the posterior pdf is usually approximated by a compact distribution centred around the \textit{maximum a posteriori} solution, typically a Gaussian pdf. This is referred to as the linearised method. In this work, we apply both nonlinear and linearised methods to 2D acoustic Bayesian FWI problems. We use a variational inference algorithm for the nonlinear case, in which a transformed Gaussian distribution is optimised to approximate the unknown, full posterior pdf. The results can then be compared with those from a linearised, locally-Gaussian based method. We also apply an independent nonlinear variational algorithm -- Stein variational gradient descent -- for comparison. The results show that while both the linearised and nonlinear methods recover the posterior mean models accurately, they exhibit significantly different posterior uncertainty structures, especially around layer interfaces, due to the linearisation of wave physics. Linearised uncertainty estimates are shown to be significantly less accurate: they provide far less accurate fits to observed waveform data, and yield biased estimates of inferred meta-properties such as volumes of geological bodies. This work therefore motivates the application of fully nonlinear inversion methods in Bayesian FWI if accurate uncertainty estimates over parameters, or inferred or interpreted meta-properties are important.
\end{abstract}

\section{Introduction}
Seismic full waveform inversion (FWI) creates high resolution images of subsurface velocity structures using seismic full waveform data recorded on the surface \cite{tarantola1984inversion, virieux2009overview}. Traditionally, FWI problems are solved using deterministic, gradient-based numerical optimisation methods. These methods identify a locally-optimal velocity model by iteratively perturbing an initial model estimate to minimise a measure of misfit between observed and simulated waveform data (seismograms) \cite{pratt1998gauss}. However, FWI solutions are highly nonunique due to imperfect acquisition geometry (typically limited to surface observations only), the existence of noise in the recorded data, nonlinearity of the forward function and the under-determined nature of real-world tomographic problems in which the target is heterogeneous over all length scales. Consequently, the optimal model solution obtained may not represent the true Earth structure with any desired level of accuracy. Estimating uncertainties in FWI solutions is therefore important for quantitative interpretation of the Earth's properties, and is crucial to enable robust risk-based decision-making based on the results \cite{arnold2018interrogation, ely2018assessing, zhao2022interrogating, siahkoohi2022deep, zhang2021interrogation}. 

Bayesian inference accounts for uncertainties by solving inverse problems probabilistically, by calculating the so-called \textit{posterior} probability density function (pdf). This is obtained by combining existing, or \textit{a priori}, knowledge about the Earth with new information provided by observed data, using Bayes' rule. The posterior pdf describes all plausible model solutions that are consistent with both the observed data and the existing prior information, and their relative probability of representing the true Earth structure.

For linear inverse problems, the posterior pdf can be calculated straightforwardly. For example, if both the likelihood function and prior pdf are described by Gaussian distributions, then the posterior pdf is also a Gaussian distribution and can be calculated analytically \cite{tarantola2005inverse}. For nonlinear inverse problems (such as FWI), the posterior pdf is not usually a Gaussian distribution. It is then common to linearise the forward function around a specific model solution, the optimal, or so-called \textit{maximum a posteriori} (MAP) solution obtained using the deterministic, iterative inversion method above. The posterior pdf around the MAP model can then be approximated by a Gaussian distribution, which accounts only for approximate, \textit{linearised} physics relating the parameters to observed data locally around that model \cite{bui2013computational, zhu2016bayesian, liu2022pre, cui2024glad, keating2026comparison}. Since accurate estimation of the MAP solution requires a good starting point for the iterative inversion, the uncertainty estimates are also strongly affected by the initial model parameter estimate. In addition, the uncertainty estimates are based on the local sensitivity of data to parameter values around the MAP model, and therefore cannot describe uncertainties arising from nonlinearity of the forward function \cite{tarantola2005inverse, galetti2015uncertainty}. In highly nonlinear problems such as FWI, this may be a serious deficiency.

Over the past few years, Bayesian FWI methods have been developed to calculate the posterior distribution within parameter space. Markov chain Monte Carlo (McMC) methods are often used to generate an ensemble of samples (model realisations) distributed approximately according to the posterior distribution \cite{mosegaard1995monte, sambridge2002monte, fichtner2019hamiltonian}. McMC methods have been applied to vairous Bayesian FWI problems \cite{biswas20172d, ray2018low, visser2019bayesian, guo2020bayesian, gebraad2020bayesian, kotsi2020time, zhao2021gradient, khoshkholgh2022full, zunino2023hmclab, de2023acoustic, berti2023computationally, dhabaria2024hamiltonian}. Statistical properties of the posterior distribution, such as the posterior mean and standard deviation values, are calculated from those samples. However, pure sampling-based methods become computationally intractable for high-dimensional inverse problems, especially in FWI problems which typically contain more than 10,000 unknown parameters, due to the curse of dimensionality \cite{curtis2001prior}.

A different class of methods called \textit{variational inference} solves nonlinear inverse problems through numerical optimisation. In variational inference, an optimal pdf is selected from a predefined family of probability distributions (referred to as the variational family) to approximate the true, unknown posterior distribution. This optimal distribution is found by iteratively minimising the difference (distance) between the variational and posterior distributions \cite{bishop2006pattern, blei2017variational, zhang2018advances}. The Kullback-Leibler (KL) divergence \cite{kullback1951information} is typically used to measure the distance between these two probability distributions. While only a small number of samples is used to estimate the KL divergence in each iteration, the optimisation succeeds stochastically by aggregating average information over samples taken in many iterations. And the constraints offered \textit{a priori} by defining a limited variational family reduce the degrees of freedom of the overall problem. As a result, variational methods have proven to be efficient, and easier to scale to high dimensional problems than pure random sampling methods \cite{nawaz2018variational, zhang2019seismic, zhao2021bayesian, siahkoohi2023reliable}, and a variety of variational methods have been applied to Bayesian FWI problems in both two and three spatial dimensions \cite{zhang2021bayesianfwi, wang2023re, zhang20233, lomas20233d, yin2024wise, zhao2024bayesian, corrales2025annealed, berti2025bayesian, chen2025variational}. 

\citet{zhao2024physically} proposed the physically structured variational inference (PSVI) algorithm for Bayesian FWI, in which a (transformed) Gaussian distribution is shaped to approximate the unknown posterior pdf. This is achieved by optimising the Gaussian mean vector and a sparse covariance matrix which reduces memory requirements and computational cost. PSVI is an efficient algorithm that has been shown to provide reasonably accurate Bayesian posterior FWI solutions in 2D \cite{zhao2024variational}, in 3D \cite{zhao2025efficient} and in time-lapse scenarios \cite{zhao2025uncertainty}. Stein variational gradient descent \cite[SVGD,][]{liu2016stein} is another variational algorithm that optimises an initial set of samples iteratively, such that the density of their final distribution approximates the posterior probability density. The final set of samples is used to estimate statistics of the posterior pdf. SVGD has also been applied to various Bayesian FWI problems \cite{zhang2021bayesianfwi, zhang20233, lomas20233d, corrales2025annealed, berti2025bayesian}, and we apply both PSVI and SVGD to problems below.

Previous studies have compared inversion results obtained from linearised and nonlinear methods in various geophysical problems. \citet{galetti2015uncertainty} performed seismic travel time tomography using both linearised and nonlinear (Monte Carlo) methods, and demonstrated that nonlinear methods offer more reasonable and intuitive uncertainty estimates that describe the features expected to arise from nonlinearity of wave physics. Specifically in results from the nonlinear inversions, loop-like uncertainty structures are observed around the boundaries of velocity anomalies, since different velocity values and shapes of velocity anomalies provide a similar fit to observed data. Such features are absent in linearised uncertainty results. In electrical resistivity tomography, \citet{galetti2018transdimensional} illustrated that including full nonlinearity enables deeper parts of the Earth to be imaged with the same data, meaning that sensitivity from nonlinear inversion extends deeper compared to linearised methods. \citet{ren2020uncertainty} performed uncertainty analysis in electromagnetic inversions using linearised and nonlinear methods, and showed that the latter explores a larger portion of parameter space than the former and hence the results are better. \citet{bloem2020experimental} compared linearised and nonlinear methods when optimising experimental designs for earthquake location surveys. Other studies performed similar comparisons in various geophysical problems \cite{fichtner2010full, sen2013global, rawlinson2014seismic, fichtner2019hamiltonian}. Nevertheless in FWI, there is currently no systematic comparison of uncertainty results between linearised and nonlinear methods.

In this paper, we perform Bayesian FWI and uncertainty quantification using both linearised and nonlinear methods. Specifically, we compare a Gaussian-based linearised method and PSVI, a nonlinear inversion method that also uses a (transformed) Gaussian kernel. The differences in the corresponding uncertainty results are principally due to the linearised and nonlinear inversion methods since everything else is held constant. To validate the inversion results obtained from PSVI, we apply SVGD as an independent nonlinear (variational) inversion method. We show that uncertainty results obtained from nonlinear methods (PSVI and SVGD) differ significantly from, and are more accurate than, those from the linearised method. An important conclusion is that the nature of uncertainty around abrupt variations in subsurface properties is quite different; we show that in turn this significantly changes derived quantities, such as geological volumetric estimates inferred from FWI solutions.

This paper is organised as follows. In Section 2, we introduce Bayesian FWI and both linearised and nonlinear inversion methods for uncertainty estimation. 
%Then we analyse how the linearised and fully nonlinear uncertainties vary when different data sets are used for inversion. 
In Section 3, we use a simple layered model with an almost ideal scenario to compare inversion results from linearised and nonlinear methods, and exemplify how differences of the results from the three methods vary with different data sets. Then we analyse a more realistic FWI example which illustrates that uncertainty estimates from nonlinear methods are more accurate than those from the linearised method. Finally, we provide a brief discussion and draw conclusions.

\section{Methodology}
\subsection{Bayesian Inference}
In Bayesian inference, the posterior probability density function (pdf) $p(\mathbf{m}|\mathbf{d}_{obs})$ of model parameters $\mathbf{m}$ given observed data $\mathbf{d}_{obs}$ is calculated using Bayes' theorem
\begin{equation}
	p(\mathbf{m}|\mathbf{d}_{obs}) = \dfrac{p(\mathbf{d}_{obs}|\mathbf{m})p(\mathbf{m})}{p(\mathbf{d}_{obs})}
	\label{eq:bayes}
\end{equation}
where $p(\mathbf{m})$ is the prior pdf of parameters $\mathbf{m}$, which describes the \textit{a priori} knowledge about $\mathbf{m}$ independently of (prior to observing) data $\mathbf{d}_{obs}$, and $p(\mathbf{d}_{obs}|\mathbf{m})$ is called the likelihood function which denotes the probability of observing $\mathbf{d}_{obs}$ given a particular model $\mathbf{m}$ (that is, under the assumption that $\mathbf{m}$ represents the true state of the Earth). Term $p(\mathbf{d}_{obs})$ is called the \textit{evidence} and here acts as a normalisation constant to ensure that $p(\mathbf{m}|\mathbf{d}_{obs})$ is a valid probability distribution.

\subsection{Linearised Method}
\label{sec:linearised}

In this study, we assume that both the prior pdf and likelihood function are defined to be Gaussian distributions
\begin{equation}
	p(\mathbf{m}) \propto \exp \left[-\frac{1}{2} (\mathbf{m} - \mathbf{m}_{prior})^T \bold\Sigma^{-1}_{prior} (\mathbf{m} - \mathbf{m}_{prior})\right]
	\label{eq:prior}
\end{equation}
\begin{equation}
	p(\mathbf{d}_{obs}|\mathbf{m}) \propto \exp \left[-\frac{1}{2} (\mathbf{d}_{syn} - \mathbf{d}_{obs})^T \bold\Sigma^{-1}_{\mathbf{d}} (\mathbf{d}_{syn} - \mathbf{d}_{obs})\right]
	\label{eq:likelihood}
\end{equation}
where $\mathbf{m}_{prior}$ is the \textit{a priori} Gaussian mean model, and $\bold\Sigma_{prior}$ is the prior covariance matrix; $\mathbf{d}_{syn} = \mathbf{F}(\mathbf{m})$ denotes synthetic data corresponding to a particular model $\mathbf{m}$, and obtained by solving a forward function $\mathbf{F}(\cdot)$, and $\bold\Sigma_{\mathbf{d}}$ is the data covariance matrix that quantifies data uncertainties. By substituting equations \ref{eq:prior} and \ref{eq:likelihood} into equation \ref{eq:bayes}, the posterior distribution becomes
\begin{equation}
	p(\mathbf{m}|\mathbf{d}_{obs}) \propto \exp (-\chi)
	\label{eq:posterior}
\end{equation}
where
\begin{equation}
	\chi = \frac{1}{2} (\mathbf{d}_{syn} - \mathbf{d}_{obs})^T \bold\Sigma^{-1}_{\mathbf{d}} (\mathbf{d}_{syn} - \mathbf{d}_{obs}) + \frac{1}{2} (\mathbf{m} - \mathbf{m}_{prior})^T \bold\Sigma^{-1}_{prior} (\mathbf{m} - \mathbf{m}_{prior})
	\label{eq:misfit}
\end{equation}
is a function of $\mathbf{m}$ since $\mathbf{d}_{syn} = \mathbf{F}(\mathbf{m})$ and everything else is fixed, and can be regarded as a misfit function between the observed and synthetic data, regularised by the Gaussian prior information \cite{asnaashari2013regularized}.

For nonlinear inverse problems (i.e., in which $\mathbf{F}(\mathbf{m})$ is nonlinear), the posterior pdf is rarely a Gaussian distribution even if both the prior pdf and likelihood function are Gaussian. Nevertheless, we can linearise the forward function in the vicinity of any model solution, in particular the \textit{maximum a posteriori} (MAP) solution. The posterior pdf can then be approximated locally around the MAP model $\mathbf{m}_{MAP}$ by a Gaussian distribution \cite{tarantola2005inverse}:
\begin{equation}
	p(\mathbf{m}|\mathbf{d}_{obs}) \propto \exp \left[-\frac{1}{2} (\mathbf{m} - \mathbf{m}_{MAP})^T \bold\Sigma^{-1}_{post} (\mathbf{m} - \mathbf{m}_{MAP})\right]
	\label{eq:posterior_gaussian}
\end{equation}
where $\mathbf{m}_{MAP}$ is often obtained by maximising equation \ref{eq:misfit} (hence the name) using an iterated, gradient-based, deterministic inversion method \cite{virieux2009overview}. $\bold\Sigma_{post}$ is the approximate posterior covariance matrix given by
\begin{equation}
	\bold\Sigma_{post} = (\mathbf{H} + \bold\Sigma_{prior}^{-1})^{-1} = \bold\Sigma_{prior}^{1/2}(\bold\Sigma_{prior}^{1/2} \mathbf{H} \bold\Sigma_{prior}^{1/2} + \mathbf{I})^{-1}\bold\Sigma_{prior}^{1/2}
	\label{eq:post_cov}
\end{equation}
where $\mathbf{I}$ is the identity matrix, $\mathbf{H} = \mathbf{J}^T \bold\Sigma^{-1}_{\mathbf{d}} \mathbf{J}$ is the Gauss-Newton approximation of the data misfit (the first term on the right hand side of equation \ref{eq:misfit}) Hessian matrix \cite{pratt1998gauss, tarantola2005inverse}, and $\mathbf{J} = {\partial \mathbf{F}}/{\partial \mathbf{m}}$ is the matrix of Fr\'echet derivatives of the forward function.

Direct calculation of the Hessian matrix (or its inverse) can be expensive in FWI due to the computational cost of forward simulation (i.e., solving a wave equation) and the dimensionality of $\mathbf{m}$ \cite{fichtner2011hessian}. Randomised singular value decomposition (SVD) algorithm can be used to estimate the Hessian matrix approximately \cite[e.g., ][]{liberty2007randomized, bui2013computational, zhu2016bayesian}. \citet{keating2026comparison} compared the performance of the algorithm to approximate the Hessian and inverse Hessian matrices respectively in FWI problems and concluded that the former results in more accurate results. In this work, we approximate the prior-conditioned Hessian matrix
\begin{equation}
	\bold\Sigma_{prior}^{1/2} \mathbf{H} \bold\Sigma_{prior}^{1/2} \approx \mathbf{V}_r \mathbf{\Lambda}_r \mathbf{V}_r^T
	\label{eq:prior_condition_Hessian}
\end{equation}
where $\mathbf{\Lambda}_r = \text{diag}(\lambda_1, ..., \lambda_r)$ is a diagonal matrix with diagonal elements equal to the dominant $i$th singular values $\lambda_i$ of the prior-conditioned Hessian matrix, and $\mathbf{V}_r$ is a matrix composed of the associated singular vectors. The posterior covariance matrix can then be approximated as
\begin{equation}
	\bold\Sigma_{post} \approx \bold\Sigma_{prior} - \hat{\mathbf{V}}_r \mathbf{D}_r \hat{\mathbf{V}}_r^T
	\label{eq:post_cov_approx}
\end{equation}
where $\hat{\mathbf{V}}_r = \bold\Sigma_{prior}^{1/2} \mathbf{V}_r$, and $\mathbf{D}_r = \text{diag}(\lambda_1/(\lambda_1+1), ..., \lambda_r/(\lambda_r+1))$ is a diagonal matrix. 

Once we have computed the MAP model $\mathbf{m}_{MAP}$ and the approximate covariance matrix $\bold\Sigma_{post}$, we can calculate statistical properties of the Gaussian posterior distribution and generate posterior samples from it \cite{bui2013computational, zhu2016bayesian}. For example, the posterior standard deviation values can be calculated by the square roots of the diagonal elements of $\bold\Sigma_{post}$; posterior samples can be obtained by
\begin{equation}
	\mathbf{m}_{post} = \mathbf{m}_{MAP} + \bold\Sigma_{post}^{1/2}\mathbf{\boldsymbol\eta}
	\label{eq:posterior_sample}
\end{equation}
where $\boldsymbol\eta$ represents random samples drawn from a multivariate standard normal distribution $\mathcal{N}(\mathbf{0}, \mathbf{I})$ of the same dimensionality as $\mathbf{m}$. Matrix $\bold\Sigma_{post}^{1/2}$ is calculated by
\begin{equation}
	\bold\Sigma_{post}^{1/2} = \bold\Sigma_{prior}^{1/2}(\mathbf{V}_r\mathbf{P}_r\mathbf{V}_r^T + \mathbf{I})
	\label{eq:sigma_half}
\end{equation}
with $\mathbf{P}_r = \text{diag}(1/\sqrt{\lambda_1 + 1} - 1, ..., 1/\sqrt{\lambda_r + 1} - 1)$ being a diagonal matrix.

Note that since the posterior distribution in equation \ref{eq:posterior_gaussian} is obtained by linearising the forward function around the MAP solution, it can not describe uncertainties caused by the nonlinearity of forward function $\mathbf{F}$.

\subsection{Nonlinear Method}
Instead of linearising the forward function around the MAP solution, fully nonlinear inversion methods have been developed and applied to FWI problems. These methods can be divided into two main types: Monte Carlo sampling \cite{gebraad2020bayesian, zhao2021gradient, biswas2022transdimensional, khoshkholgh2022full} and variational inference \cite{zhang2021bayesianfwi, izzatullah2023physics, zhao2024bayesian, yin2024wise}. Since sampling-based methods generally suffer from slow convergence due to the curse of dimensionality \cite{curtis2001prior}, in this work we use variational inference to calculate an approximate posterior pdf within the parameter space to quantify uncertainties. 

In variational inference, we define a family of probability distributions $\mathcal{Q} = \{q(\mathbf{m})\}$, from which we select an optimal member $q^*(\mathbf{m})$ to best approximate the true posterior pdf. This optimal distribution can be found by minimising the Kullback-Leibler (KL) divergence \cite{kullback1951information} between the variational distribution $q(\mathbf{m})$ and the posterior distribution $p(\mathbf{m}|\mathbf{d}_{obs})$
\begin{equation}
	\text{KL}[q(\mathbf{m})||p(\mathbf{m}|\mathbf{d}_{obs})] = \mathbb{E}_{q(\mathbf{m})}[\log q(\mathbf{m}) - \log p(\mathbf{m}|\mathbf{d}_{obs})]
	\label{eq:kl}
\end{equation}
The KL divergence of two distributions is non-negative and equals zero when the two distributions are identical. Minimising the KL-divergence is mathematically equivalent to maximising an evidence lower bound (ELBO) of $\log p(\mathbf{d}_{obs})$:
\begin{equation}
	\text{ELBO}[q(\mathbf{m})] = \mathbb{E}_{q(\mathbf{m})}[\log p(\mathbf{m}, \mathbf{d}_{obs}) - \log q(\mathbf{m})]
	\label{eq:elbo}
\end{equation}
This can be solved using gradient-based optimisation algorithms, and the optimisation result -- the best posterior approximation $q^*(\mathbf{m})$ -- is fully probabilistic.

We use a physically structured variational inference algorithm \cite[PSVI,][]{zhao2024physically}, which adapts a Gaussian distribution with physics-informed covariances to approximate the posterior distribution. Given that a Gaussian distribution is defined over the space of real numbers and that in most geophysical imaging problems model parameters are bounded by physical constraints (e.g., seismic velocity should be a positive number), an invertible transform (a bijection) is applied to the Gaussian random variables to ensure that the transformed model parameters satisfy their physical constrains. We use the commonly used \textit{logit} functions \cite{kucukelbir2017automatic, zhang2019seismic}
\begin{equation}
	\begin{split}
		& m_i = f(\theta_i) = a_i + \dfrac{b_i - a_i}{1+\exp (-\theta_i)}\\
		& \theta_i = f^{-1}(m_i) = \log(m_i - a_i) - \log(b_i - m_i) 
	\end{split}
	\label{eq:log_transform}
\end{equation}
where $\theta_i$ is a Gaussian random variable defined in an unconstrained space (from minus to plus infinity), and $m_i$ is the converted model parameter bounded by the lower and upper bounds $a_i$ and $b_i$, respectively. We then optimise the Gaussian variational distribution within the unconstrained space. Note that values of $a_i$ and $b_i$ are fixed during inversion.

Optimising a full Gaussian covariance matrix is extremely expensive in FWI problems in terms of both memory requirements and computational cost, since the parameter space dimensionality of FWI is typically higher than 10,000. PSVI constructs a sparse covariance matrix by modelling only the most important correlations in model vector $\mathbf{m}$. In imaging inverse problems, this approach focuses on posterior correlations between pairs of parameters that are in spatial proximity to each other, typically within one dominant wavelength in FWI; all other parameter correlations are set to zero.

This manifestation of PSVI has been demonstrated to be an efficient variational method that provides reasonable estimates of statistics of the full, nonlinear posterior distribution \cite{zhao2025efficient, zhao2025bayesian}. More importantly, both PSVI and the linearised method described in Section \ref{sec:linearised} use Gaussian distributions to approximate the posterior pdf. The main difference between the two algorithms is that the former seeks an optimal (transformed) Gaussian distribution that minimises the KL-divergence (equation \ref{eq:kl}) within full parameter space, whereas the latter introduces a Gaussian distribution that represents only local covariances around the MAP model. The former therefore accounts for nonlinear uncertainties whereas the latter only captures locally linearised uncertainties, and analysing the inversion results allows us to compare the two methods in FWI problems.

Given that there is no closed-form Bayesian posterior solution to most nonlinear inverse problems so it is never clear whether estimates of Bayesian uncertainties are accurate, we employ a completely independent method -- Stein variational gradient descent \cite[SVGD,][]{liu2016stein} -- to cross-validate the posterior solutions obtained from PSVI and the linearised method. SVGD is an effective variational inference algorithm that iteratively updates a set of model samples generated from an initial distribution (e.g., the prior distribution) into a set of samples from the corresponding posterior distribution, by minimising the KL-divergence between the posterior pdf and the samples' density \cite{liu2016kernelized}. Importantly for this study, the SVGD solution is not constrained to follow a Gaussian-like distribution (or any other particular form). It therefore provides an independent assessment of the true, non-linearised uncertainty.

Finally note that the nonlinearity of an inverse problem often depends on the experimental design used to collect the data \cite{curtis2004theory2}. Therefore, differences between linearised and nonlinear inversions can also change with the design. We will show that there are three broad classes of such differences. First, if we have a design that produces highly informative data such that the posterior distribution is unimodal and very tightly constrained around a single solution, then, uncertainty estimates from both methods might be similar. Second, if data are barely sensitive to model parameters then posterior uncertainties will be dominated by the common prior distribution, results from both methods will also be similar. However, in the third class in which information in the posterior solution is dominated by neither the data nor the prior distribution, the posterior pdf will depend on the nonlinearity of the problem and we would observe different posterior statistics from linearised and nonlinear methods.

\section{Results}
We now apply the single linearised and two nonlinear (PSVI and SVGD) inversion methods to 2D, constant-density, acoustic FWI problems. For all three methods, we use the same Gaussian distributions to define the prior pdf and likelihood function, such that the posterior solutions obtained are directly comparable. The nonlinear forward problem involves solving a 2D scalar wave equation, which is performed using a time domain finite difference method \cite{moczo2007finite}. Gradients of the logarithmic likelihood function (equation \ref{eq:likelihood}) with respect to velocities are computed using the adjoint state method \cite{plessix2006review, fichtner2006adjoint}.

\subsection{Layered Model}
\begin{figure}
	\centering\includegraphics[width=\textwidth]{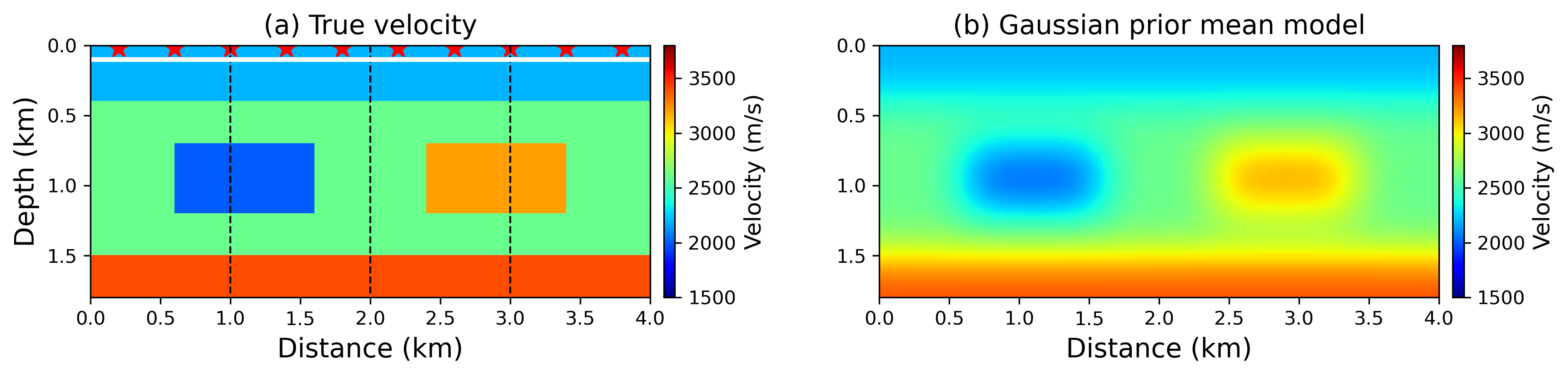}
	\caption{(a) True P-wave velocity used in the layered model example. Source locations are indicated by red stars and a receiver line is marked by a white line. Dashed black lines display locations of three vertical profiles used to compare the inversion results in the main text. (b) Velocity model used to define the mean of a Gaussian prior distribution, obtained by smoothing the true velocity model in (a).}
	\label{fig:layered_true_initial}
\end{figure}

\begin{figure}
	\centering\includegraphics[width=0.8\textwidth]{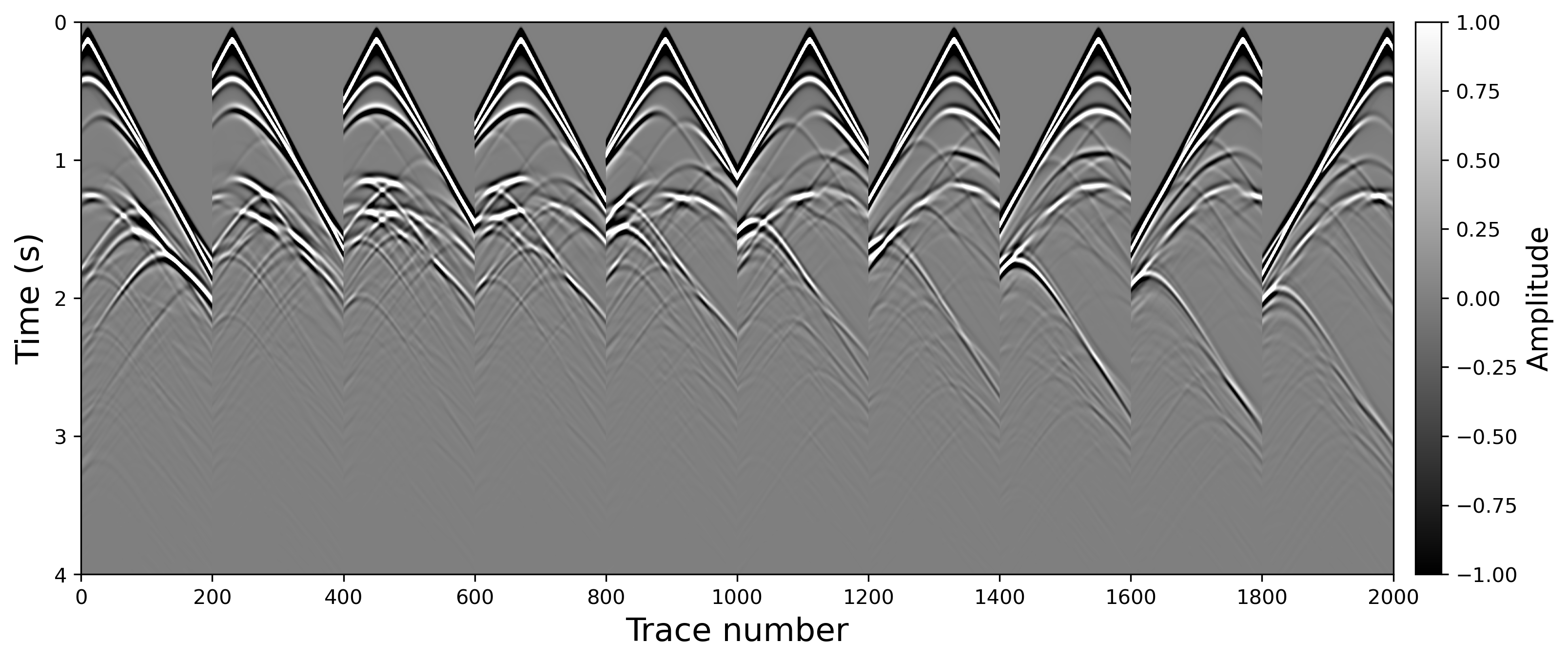}
	\caption{Ten common shot gathers with a dominant frequency of 10 Hz, used as the observed waveform data in the first FWI example.}
	\label{fig:layered_observed_data}
\end{figure}

The first example defines the true structure to consist of a simple layered model with a low- and a high-velocity rectangular anomaly in the second layer, as shown in Figure \ref{fig:layered_true_initial}a. This simple velocity model enables us to analyse the inversion results from different methods intuitively. The model is discretised into a regular grid of 90 $\times$ 200 cells with a cell size of 20 m in both directions. Ten explosive sources are located at 20 m depth with a spacing of 400 m (red stars in Figure \ref{fig:layered_true_initial}a), and a receiver line containing 200 equally spaced receivers is placed at a depth of 100 m (white line in Figure \ref{fig:layered_true_initial}a). Observed waveform data are generated by solving a 2D acoustic wave equation using a Ricker wavelet with a 10 Hz dominant frequency, and are displayed in Figure \ref{fig:layered_observed_data}.

We smooth the true velocity model, and the resultant model in Figure \ref{fig:layered_true_initial}b is used to define the mean of the Gaussian prior model (i.e., $\mathbf{m}_{prior}$ in equation \ref{eq:prior}). We define a diagonal prior covariance matrix $\bold\Sigma_{prior}$ with a value of 300 m/s for all diagonal elements. Velocities above the receiver line are fixed at their true values during inversion. A diagonal data covariance matrix $\bold\Sigma_{\mathbf{d}}$ with a diagonal value of 0.1 is defined for the likelihood function (equation \ref{eq:likelihood}), which ignores correlations among waveform data points. Note that in this simple example, we deliberately choose a reasonably good prior mean model (Figure \ref{fig:layered_true_initial}b) and do not add observational noise to the data. These settings ensure that each method converges to a posterior solution that is in some sense close to the true structure in this FWI problem, allowing us to analyse and compare the uncertainty results obtained from different methods more directly (without the solution being biased by convergence to local misfit minima). In the next example, we perform FWI under more realistic conditions.

\subsubsection{\textbf{Inversion Results}}
\label{sec:layer_inversult_results}

\begin{figure}
	\centering\includegraphics[width=\textwidth]{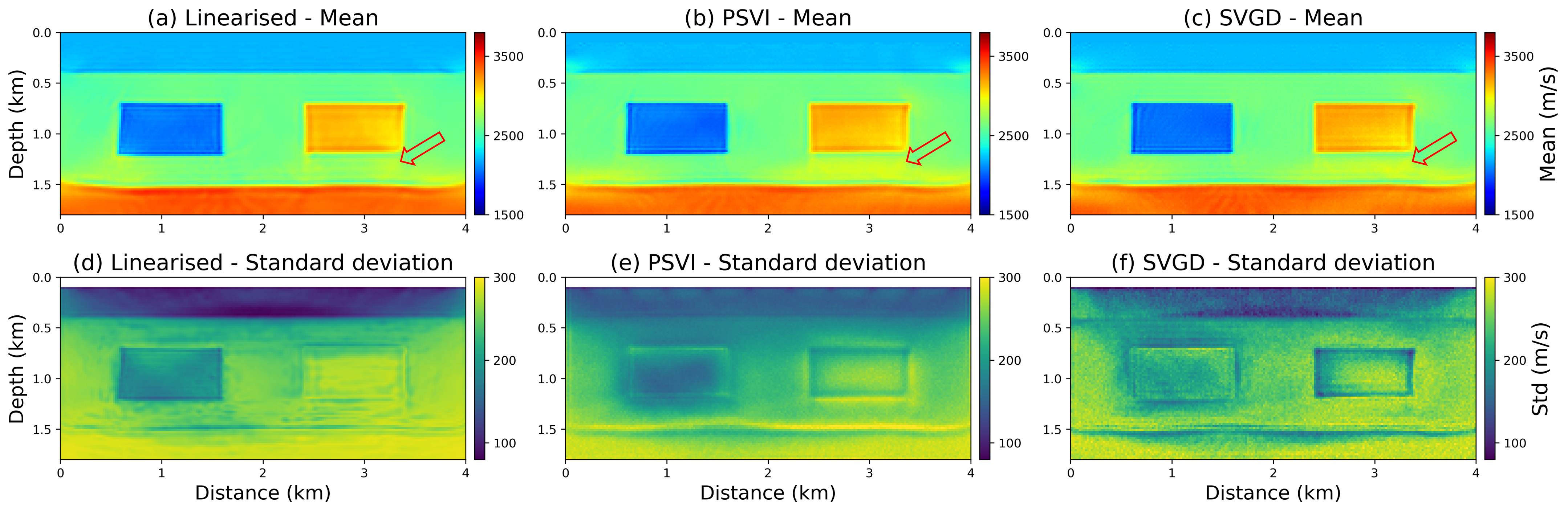}
	\caption{Inversion results obtained using linearised, PSVI and SVGD methods. (a), (b) and (c) Posterior mean models. The mean model in (a) represents the \textit{maximum a posteriori} (MAP) solution obtained by a deterministic, linearised inversion. (d), (e) and (f) The corresponding standard deviation maps. Note that velocity values above the receiver line (100 m depth) are fixed at their true values. The corresponding regions in the standard deviation maps are therefore left blank.}
	\label{fig:layered_mean_std}
\end{figure}

For the linearised inversion, we first estimate the \textit{maximum a posteriori} (MAP) model $\mathbf{m}_{MAP}$ by minimising the regularised misfit function in equation \ref{eq:misfit} using the L-BFGS algorithm \cite{liu1989limited}. To avoid local minima, we adopt a multiscale strategy \cite{bunks1995multiscale} by first inverting for a background velocity model using low frequency waveform data with a dominant frequency of 4 Hz. The initial model for this inversion is set to be the prior mean model displayed in Figure \ref{fig:layered_true_initial}b. The resulting velocity model is then used as the starting point for the inversion of 10 Hz waveform data. Figure \ref{fig:layered_mean_std}a shows the final velocity map, which for linearised inversions represents both the posterior MAP $\mathbf{m}_{MAP}$ solution and the mean of the Gaussian posterior distribution. 

We then use the randomised singular value decomposition (SVD) algorithm introduced in Section \ref{sec:linearised} (equations \ref{eq:post_cov}, \ref{eq:prior_condition_Hessian} and \ref{eq:post_cov_approx}) to estimate the Gaussian posterior covariance matrix $\bold\Sigma_{post}$, which describes linearised uncertainties around the MAP solution. To estimate the Hessian matrix accurately using the randomised SVD algorithm, we use a large number of random vectors (17,000 -- matching the dimensionality of this FWI problem) to sample it. Figure \ref{fig:layered_mean_std}d shows the square roots of the diagonal elements of the estimate of $\bold\Sigma_{post}$, representing the posterior standard deviation values. Note that parameter values above the receiver line (100 m depth) are fixed at their true values; we therefore leave the corresponding regions in the standard deviation map blank.

For both PSVI and SVGD, we invert the 10 Hz waveform data directly, since both algorithms are supposed to find reasonable posterior solutions regardless of the initial values used in this simple, nearly ideal inversion experiment. A \textit{probabilistic} multiscale strategy is shown to improve the inversion results in more strongly nonlinear FWI problems by \citet{zhang2021bayesianfwi}, \citet{lomas20233d} and \citet{zhao2025efficient} -- which will be applied in the next section for the inversion of higher frequency (17 Hz dominant frequency) data. Figures \ref{fig:layered_mean_std}b and c show the posterior mean velocity maps obtained using PSVI and SVGD, respectively, and Figures \ref{fig:layered_mean_std}e and f display the corresponding standard deviation maps.

\begin{figure}
	\centering\includegraphics[width=\textwidth]{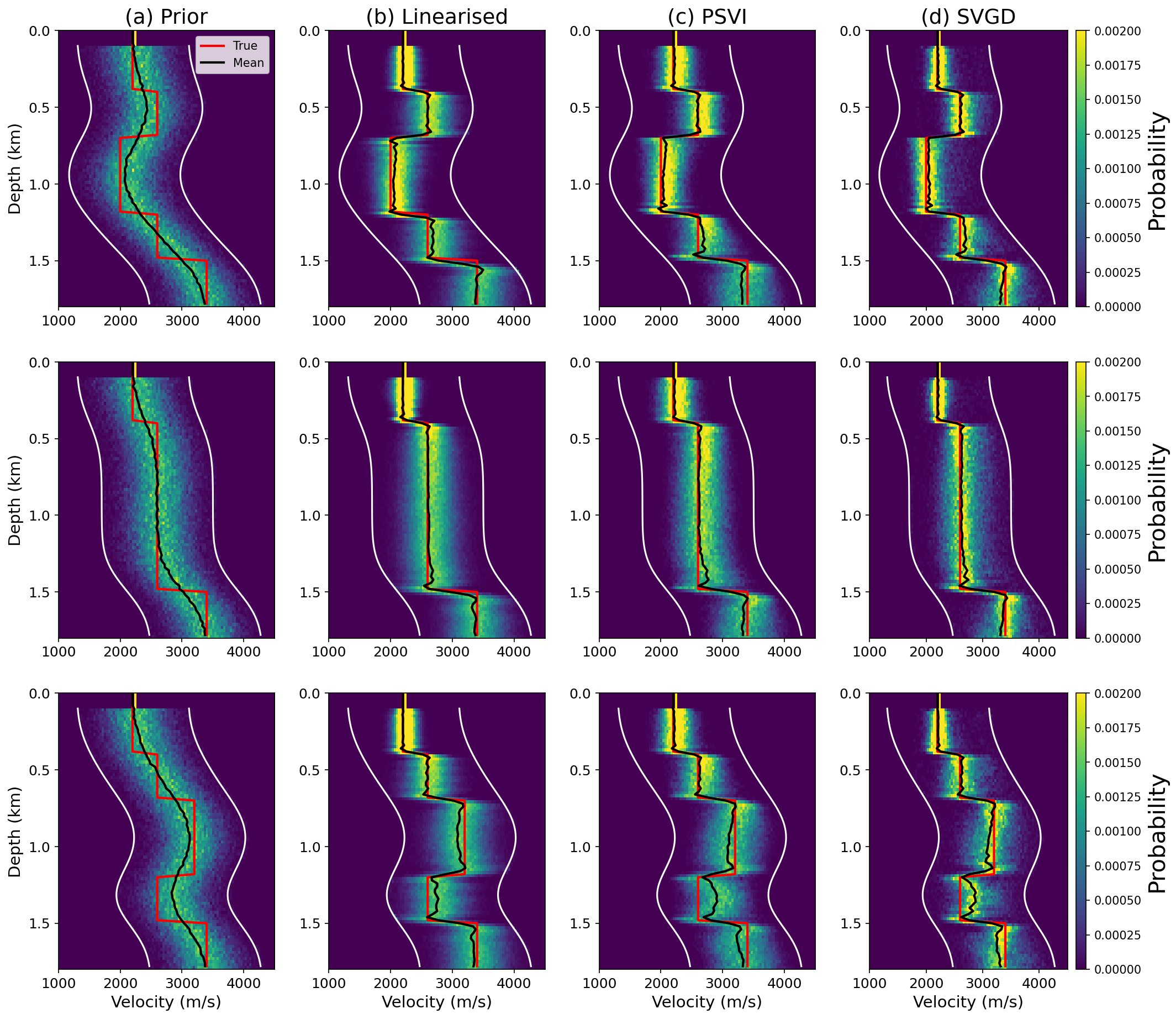}
	%\centering\includegraphics[width=\textwidth]{layered_mean_trace_vertical_10hz.png}
	\caption{(a) Prior and posterior marginal distributions obtained from (b) linearised, (c) PSVI and (d) SVGD, along three vertical profiles at horizontal locations 1 km, 2 km and 3 km. Their locations are represented by vertical dashed black lines in Figure \ref{fig:layered_true_initial}a. From top to bottom, each row represents results along one profile.}
	\label{fig:layered_traces}
\end{figure}

The three mean velocity maps exhibit similar features across most locations, generally resembling the true velocity map in Figure \ref{fig:layered_true_initial}a. To better compare the inversion results, in Figure \ref{fig:layered_traces} we display prior and posterior pdf's of three vertical velocity profiles at horizontal locations 1 km, 2 km and 3 km, respectively. These profiles correspond to locations within the low velocity anomaly, background velocity region and the high velocity anomaly in the second layer, as marked by three dashed black lines in Figure \ref{fig:layered_true_initial}a. Each row shows the results from one vertical profile. In each panel, a red line represents the true velocity profile; a black line in Figure \ref{fig:layered_traces}a shows the prior mean profile from Figure \ref{fig:layered_true_initial}b, and that in Figures \ref{fig:layered_traces}b, \ref{fig:layered_traces}c and \ref{fig:layered_traces}d shows the posterior mean profile from the corresponding method. Generally, the three sets of results recover the true velocity values in each layer. Some discrepancies between the true velocity and the inversion results are observed, such as in the layer between 1.2 km and 1.5 km depth in the bottom row in Figure \ref{fig:layered_traces}, which is also annotated by three red arrows in Figures \ref{fig:layered_mean_std}a, b and c. Possible reasons for the inaccuracy include (1) the prior mean model is inaccurate within this layer (in Figure \ref{fig:layered_traces}a); and (2) the challenge in accurately recovering a low velocity layer between two relatively higher velocity layers in FWI \cite{virieux2009overview, warner2016adaptive}, especially in deeper part of the model where waveform data have low sensitivity. In the bottom row, the posterior mean profile from the linearised method is more accurate than those from the two nonlinear methods, since the multiscale strategy for the linearised inversion recovers a more accurate long wavenumber (background) mean velocity model. Note that in nonlinear methods, the mean model is merely a posterior statistic which is not inverted directly from observed data, and is not necessarily a valid model sample that fits the data well \cite[e.g., see][]{zhang2021bayesianfwi}. Nevertheless, the true velocity profile is included within the posterior distributions obtained from the three methods.

At first glance, main features from the three standard deviation maps in Figure \ref{fig:layered_mean_std} are similar. For example, uncertainties in first layer are lower than those in the deeper layers, since waveform data have the highest sensitivity in this layer. We also observe respectively high and low uncertainties inside of the high and low velocity anomalies (away from layer boundaries), across all three panels. Since seismic wavelength is longer in a high velocity than in a low velocity body and since the same grid cell size is used throughout, the high velocity anomaly is sampled by fewer wavelengths than the low velocity anomaly; velocity values of cells within the high velocity body are thus less well constrained and can vary more while still fitting the observed data, compared to those within the low velocity body. This represents the well-known trade-off between model resolution and model uncertainty \cite{backus1968resolving, latallerie2025towards}.

However, details in the standard deviation map in the left panel differ from those in the right two panels, especially at layer boundaries where velocity values change. To demonstrate, we calculate the Structure Similarity Index Model (SSIM) between each two of the three standard deviation maps in Figure \ref{fig:layered_mean_std}. SSIM is a common measure to quantify the similarity between two images, and higher values indicate higher similarity \cite{levy2022using}. SSIM between Figures \ref{fig:layered_mean_std}d \& e, d \& f, and e \& f are 0.158, 0.180 and 0.358, respectively, meaning that the two nonlinear uncertainty maps are more similar and consistent. We then calculate the following quantity:
\begin{equation}
	R_\sigma = \dfrac{\left|2\sigma_{Linear} - \sigma_{PSVI} - \sigma_{SVGD}\right|}{2\left|\sigma_{PSVI} - \sigma_{SVGD}\right|}
	\label{eq:relative_deviation}
\end{equation}
where $\sigma_{*}$ denotes a standard deviation map from a particular method represented by the subscript. $R_\sigma$ describes relative deviation of the linearised uncertainty map from the two nonlinear ones. Generally, $R_\sigma$ values lower than 1 indicate that the linearised uncertainty estimates are roughly as good as the nonlinear ones, given the epistemic uncertainties over which nonlinear method is the better (quantified by the denominator); otherwise, the linearised estimates significantly are different from the nonlinear ones. 

\begin{figure}
	\centering\includegraphics[width=0.5\textwidth]{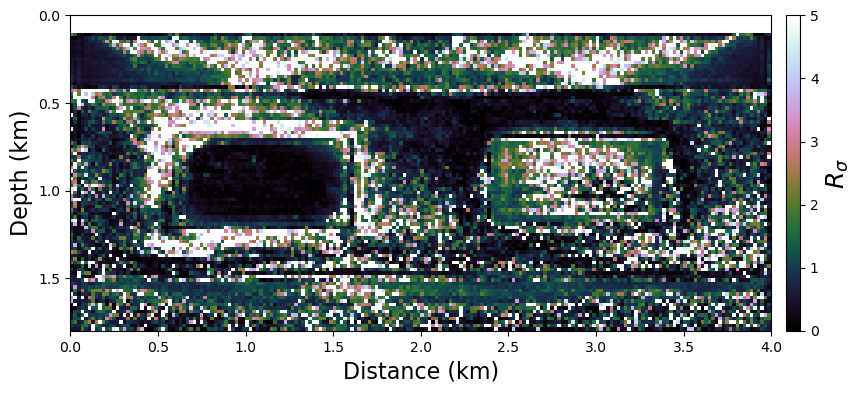}
	\caption{Relative deviation of the linearised uncertainty estimates from the nonlinear estimates ($R_\sigma$) calculated by equation \ref{eq:relative_deviation}. $R_\sigma$ values lower than 1 indicate that the linearised uncertainty estimates are roughly as good as the nonlinear ones, given the epistemic uncertainties around the various methods used herein; otherwise, the linearised estimates are different from (worse than) the nonlinear ones.}
	\label{fig:layered_r_srd}
\end{figure}

Figure \ref{fig:layered_r_srd} displays the results. We roughly observe larger $R_\sigma$ values around layer interfaces, meaning that uncertainty estimates from the linearised and nonlinear methods are mainly different around layer boundaries. There are consistently high values across most of the interior of the shallow layer down to 0.5 km. We observe clear, high $R_\sigma$ values in a loop surrounding the low velocity anomaly on the left, and low values on its interior. This might suggest that linearised methods might be expected to provide relatively misleading constraints on the boundary and hence on the size of low velocity anomalies. By contrast, there are low values surrounding the high velocity anomaly on the right, and high values throughout its interior, suggesting that the posterior distribution from linear methods might constrain the size of the high velocity region roughly as well as the nonlinear methods in this example.

Particularly, in Figures \ref{fig:layered_mean_std}e and f we observe higher uncertainty structures around layer interfaces. For example, high uncertainty loops are present around the boundaries of the two velocity anomalies. These structures are absent in the linearised inversion results in Figure \ref{fig:layered_mean_std}d. They arise since model samples with different anomaly shapes and velocity values can fit the observed waveform data to within their uncertainties almost equally well. \citet{galetti2015uncertainty} observed similar loop-like uncertainty structures in travel time tomographic problems using nonlinear (Monte Carlo) inversion methods, and attributed them to the nonlinearity (second- and higher-order aspects) of ray path physics. Here we show the existence of the uncertainty loops in reflection FWI using fully nonlinear inversion methods. These loops are clear in Figure \ref{fig:layered_mean_std}e from PSVI and exist but are less obvious in Figure \ref{fig:layered_mean_std}f from SVGD.

Around anomaly boundaries in both Figures \ref{fig:layered_mean_std}e and f, high-uncertainty loops are located inside of the low velocity anomaly and outside of the high velocity anomaly; relatively low-uncertainty loops are observed on the opposite sides of each anomaly. In other words, close to a velocity interface, higher uncertainty values typically appear on the low velocity side (an exception at the interface between the first and second layers at 0.4 km depth occurs because the reflection waveform data are highly sensitive to the top most layer as discussed above). Amplitude information in waveform data is more sensitive to subsurface impedance contrasts, and thus also to layer boundaries \cite[e.g., ][]{shuey1985simplification}, so we observe lower uncertainties around layer boundaries. Given that relatively low frequency data (10 Hz) are used in this example, waves travelling on high and low velocity sides of an interface interfere with one another in the recorded seismograms (which is part of the nonlinearity of wave physics), and thus the two wave packets can not be distinguished clearly and can not constrain parameters on the low and high velocity sides independently. Along the layer interface, seismic waves travel faster on the high velocity side. Therefore in the recorded data, first arrival travel times of the interfering waves carry information mostly about parameters on the high velocity side. Travel times of later arrival waves are hidden within the interfering wave packet and can not impose enough constraints on velocity values on the low velocity side. Therefore, lower uncertainties are observed on the high velocity side. However, from the linearised uncertainty map in Figure \ref{fig:layered_mean_std}d (and in Figure \ref{fig:layered_mean_std_scatter} below), the opposite is true: high uncertainties consistently appear on the high velocity side. These differences between the uncertainty estimates from the linearised and nonlinear methods occur because the former ignores the nonlinearity of wave physics. Note, though, that this analysis is also frequency dependent. As we will show in the next example using higher frequency waveform data, the interference becomes weaker, and low uncertainties are thus not presented on the high velocity side around layer interfaces (e.g., see regions around the low velocity body in Figure \ref{fig:layered_mean_std_17hz} and in Figure \ref{fig:layered_mean_std_scatter_17hz}b).

To further support this observation, we analyse how posterior standard deviation values vary with inverted mean velocities near layer interfaces. In the true velocity model, four velocity interfaces exist: (1) the top layer interface, (2) the boundary of the low velocity anomaly, (3) the boundary of the high velocity anomaly, and (4) the bottom layer interface, respectively. Near each interface, we display the posterior mean and standard deviation values for cells located within a range of two grid cells from the boundary interface in a scatter plot, as shown in each panel in the top row in Figure \ref{fig:layered_mean_std_scatter}. Red, blue and yellow dots display the corresponding values obtained from the linearised, PSVI and SVGD inversion results. Note that we observe distinct clusters of points in Figures \ref{fig:layered_mean_std_scatter}a and d, and each such cluster represents values extracted from cells that lie a certain distance away from the considered layer interface. We also calculate the best fit line for each set of values from a specific method, and display the results in the bottom row in Figure \ref{fig:layered_mean_std_scatter}. In Figure \ref{fig:layered_mean_std_scatter}a around the top interface, the results from the three methods show a consistent trend in which the posterior uncertainties increase with the mean velocities. However, around the other three interfaces where data are less informative, the standard deviation values from the two nonlinear methods generally have a decreasing trend as the mean values increase, yet those from the linearised method have an increasing trend.

\begin{figure}
	\centering\includegraphics[width=\textwidth]{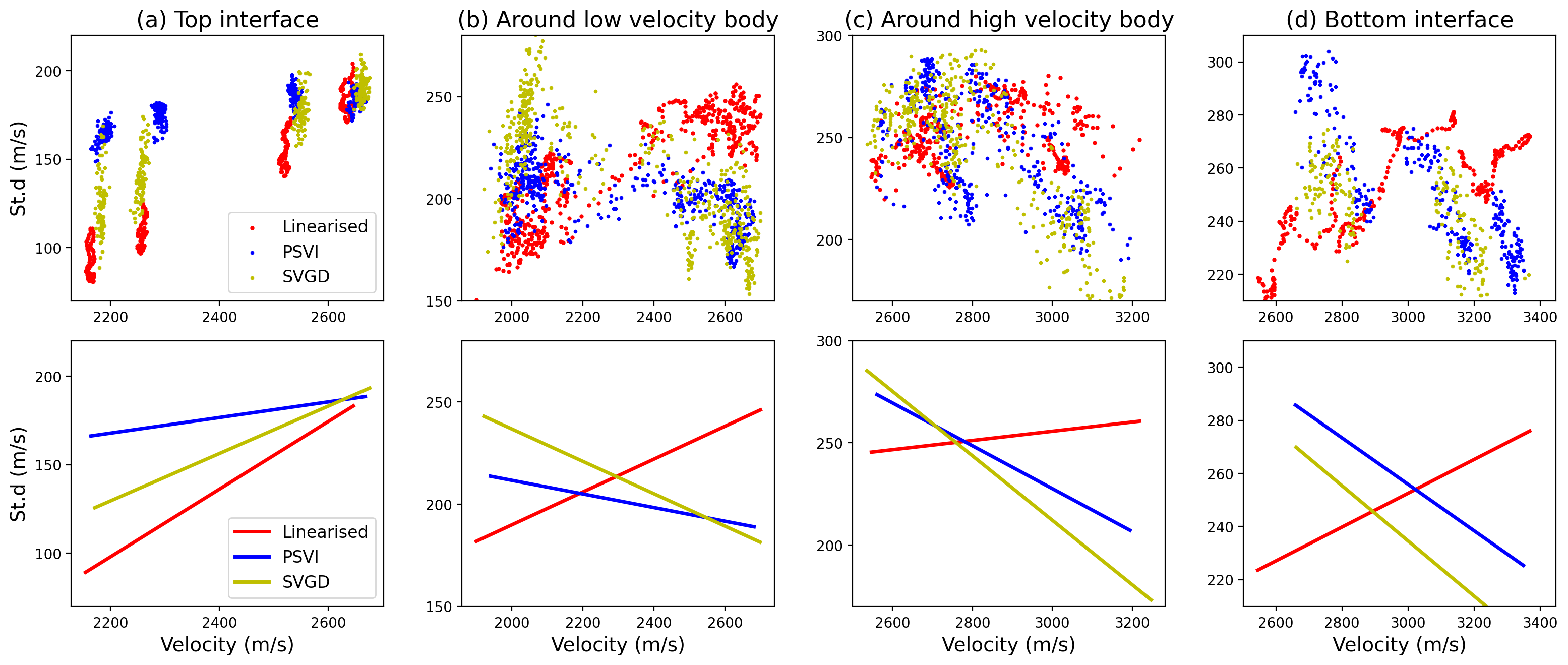}
	\caption{Top row: scatter plot showing how posterior standard deviation (St.d. in the figure) values vary with inverted mean velocities near (a) the top layer interface, (b) the boundary of the low velocity anomaly, (c) the boundary of the high velocity anomaly, and (d) the bottom layer interface. Near each interface, the posterior mean and standard deviation values for cells located within a range of two grid cells from the boundary interface are displayed. Red, blue and yellow dots display the corresponding values obtained from linearised, PSVI and SVGD methods. Bottom row: best lines that fit to the corresponding data points of different colours in the top row, showing linear trends of different results.}
	\label{fig:layered_mean_std_scatter}
\end{figure}

In summary, this simple example demonstrates that while both linearised and nonlinear inversion methods produce similar and accurate posterior mean models, different uncertainty structures are obtained when informative data are not sensitive enough to constrain model parameters well. In particular around layer interfaces, nonlinear methods tend to provide higher uncertainties on the low velocity sides and low uncertainties on the high velocity sides, whereas the linearised method exhibits an opposite pattern.

\subsubsection{\textbf{Uncertainty Results from Higher Frequency Data}}
\label{sec:uncertainty_17hz}

In this section we compare FWI uncertainty results from the linearised and nonlinear methods using more informative waveform data. Specifically, we now reproduce the FWI experiment in the previous section, but using waveform data with a dominant frequency of 17 Hz. We use a multiscale strategy for both linearised, deterministic inversion \cite{bunks1995multiscale} and probabilistic inversion \cite{zhang2021bayesianfwi, lomas20233d, zhao2025efficient}, in which results from 10 Hz data are used as the initial value for the inversion of 17 Hz data (here `value' refers to the reference model for the linearised inversion and to the initial probability distribution for the optimisation of the variational inversion). 

\begin{figure}
	\centering\includegraphics[width=\textwidth]{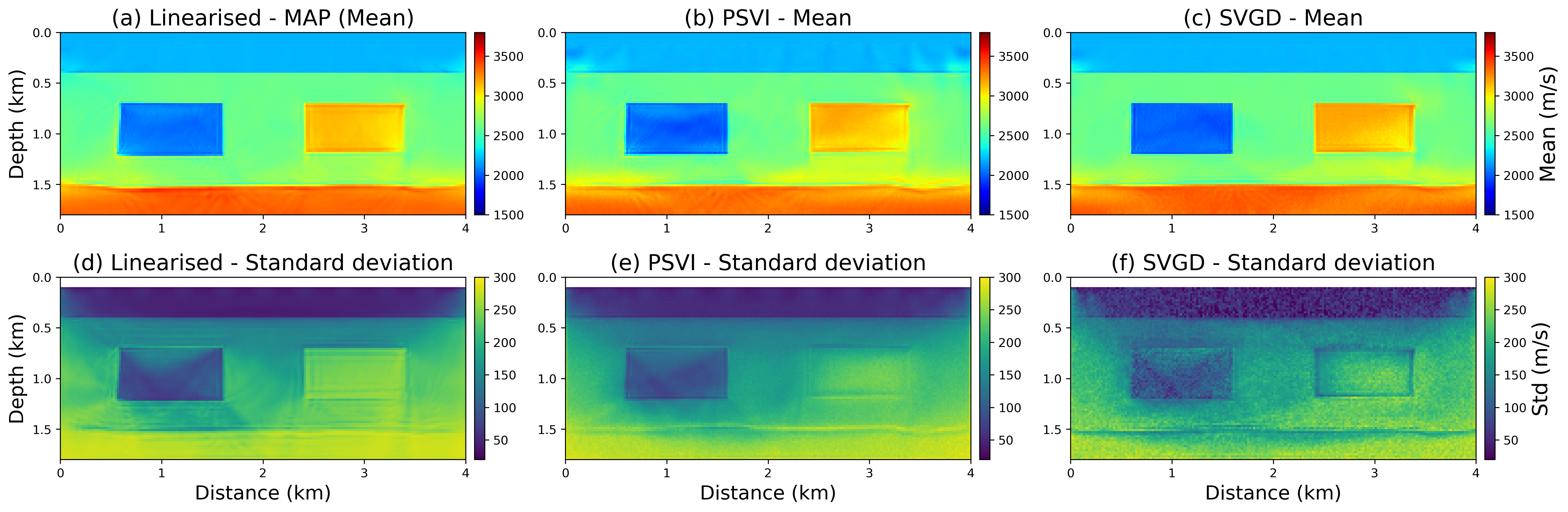}
	\caption{Inversion results using 17 Hz waveform data, obtained from (a) linearised, (b) PSVI and (c) SVGD methods. Key as in Figure \ref{fig:layered_mean_std}.}
	\label{fig:layered_mean_std_17hz}
\end{figure}

\begin{figure}
	\centering\includegraphics[width=\textwidth]{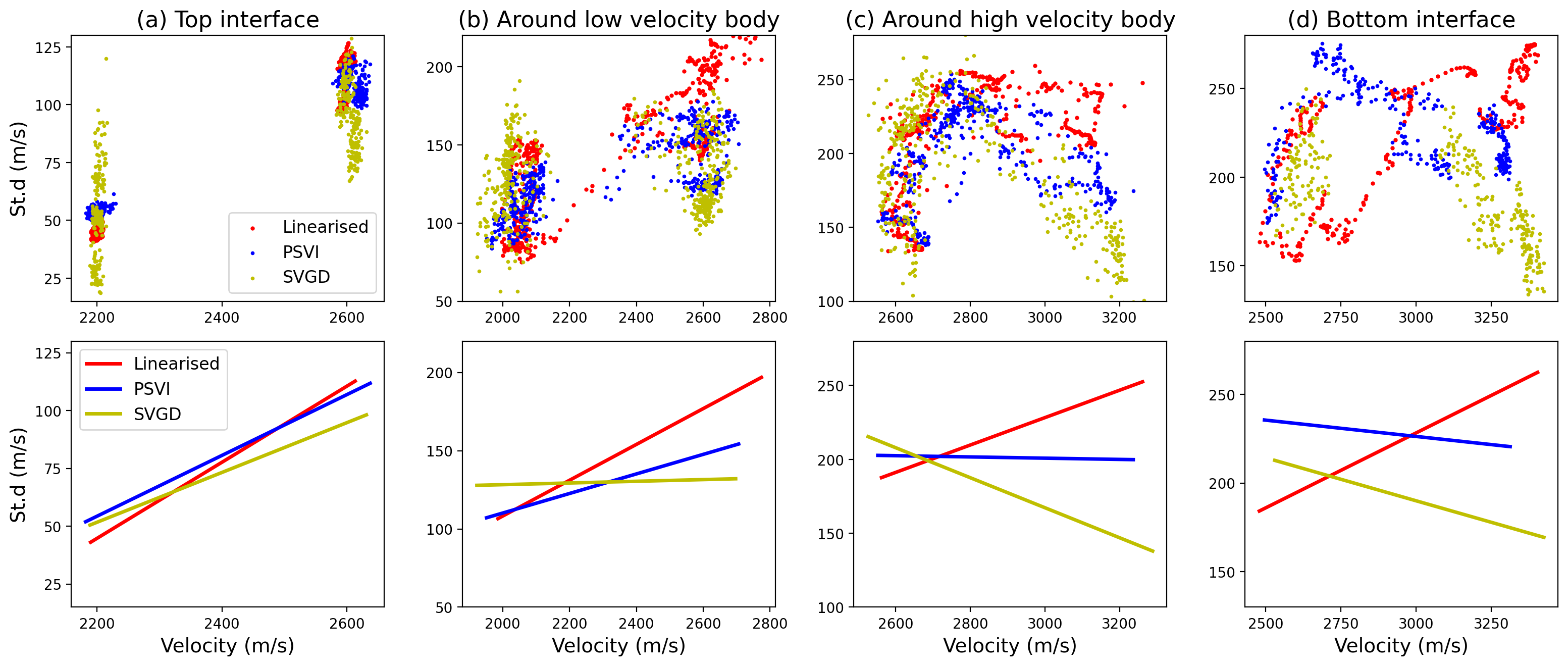}
	\caption{Scatter plots (top row) and the corresponding linear trends (bottom row) of the inversion results displayed in Figure \ref{fig:layered_mean_std_17hz} obtained using 17 Hz waveform data. Key as in Figure \ref{fig:layered_mean_std_scatter}.}
	\label{fig:layered_mean_std_scatter_17hz}
\end{figure}

Figure \ref{fig:layered_mean_std_17hz} shows the posterior mean (MAP) and standard deviation models obtained from the three methods, and Figure \ref{fig:layered_mean_std_scatter_17hz} displays the corresponding scatter plots of the posterior mean and standard deviation values around the four interfaces, similarly to those displayed in Figure \ref{fig:layered_mean_std_scatter}. Again, the three mean (MAP) velocity maps are highly consistent, all resembling the true velocity model in Figure \ref{fig:layered_true_initial}a. Since 17 Hz waveform data contain more detailed information about the subsurface compared to 10 Hz data, model parameters and their uncertainties are now better constrained, and the three standard deviation maps exhibit more similar features compared to those in Figure \ref{fig:layered_mean_std}. For example, we observe low, and almost identical uncertainty values in the top layer where waveform data have the highest sensitivity; uncertainty structures around the low velocity anomaly are consistent from the three methods. In Figures \ref{fig:layered_mean_std_scatter_17hz}a and b, the posterior standard deviation values all present a similarly increasing trend with the posterior mean velocities. The main reason for these consistent uncertainty results from linearised and nonlinear methods is that the true posterior distribution becomes (almost) unimodal and is well constrained around a single model solution given the highly informative data \cite{curtis2004theory2, strutz2023variational}. On the other hand as data become less sensitive at deeper levels, greater differences are observed between the linearised and nonlinear methods, e.g., around the bottom interface at 1.5 km depth. In Figure \ref{fig:layered_mean_std_scatter_17hz}d (particularly in the bottom panel), standard deviation values increase with velocity values from the linearised inversion results, whereas they decrease as velocities increase from the nonlinear inversion results.

\subsubsection{\textbf{Uncertainty Results from Sparse Data}}
\label{sec:uncertainty_sparse}

\begin{figure}
	\centering\includegraphics[width=\textwidth]{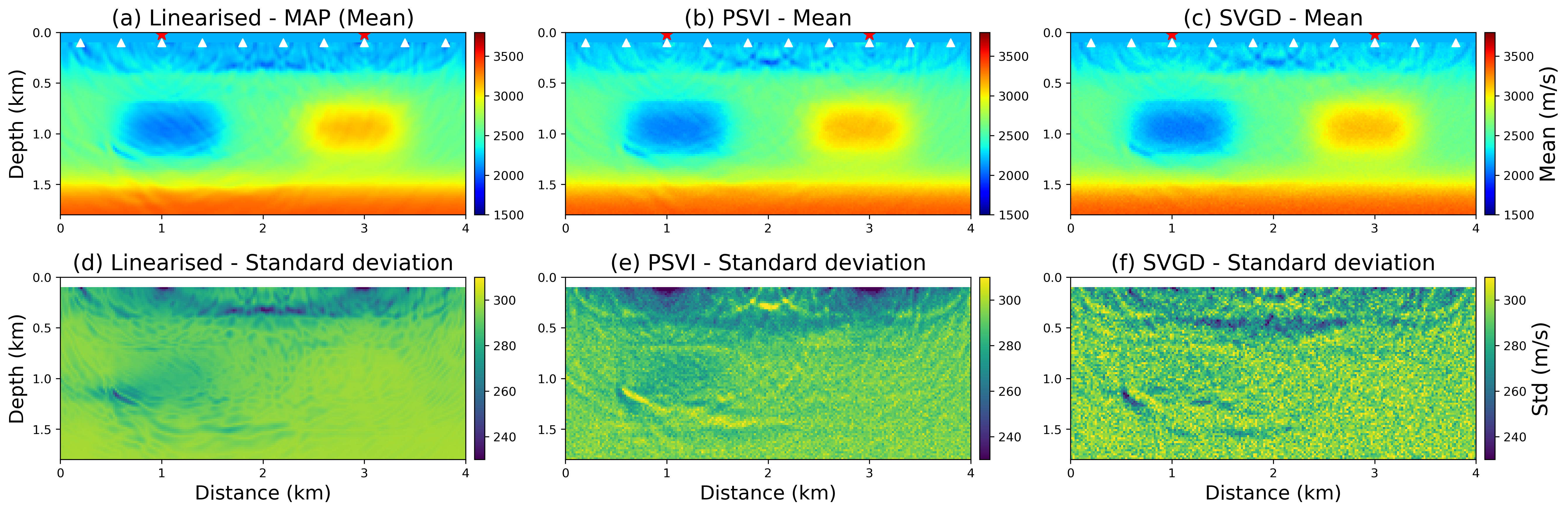}
	\caption{Inversion results using 10 Hz waveform data collected by a sparse acquisition geometry. Red stars and white triangles represent source and receiver locations, respectively. Key as in Figure \ref{fig:layered_mean_std}.}
	\label{fig:layered_mean_std_10hz_s2r10}
\end{figure}

\begin{figure}
	\centering\includegraphics[width=\textwidth]{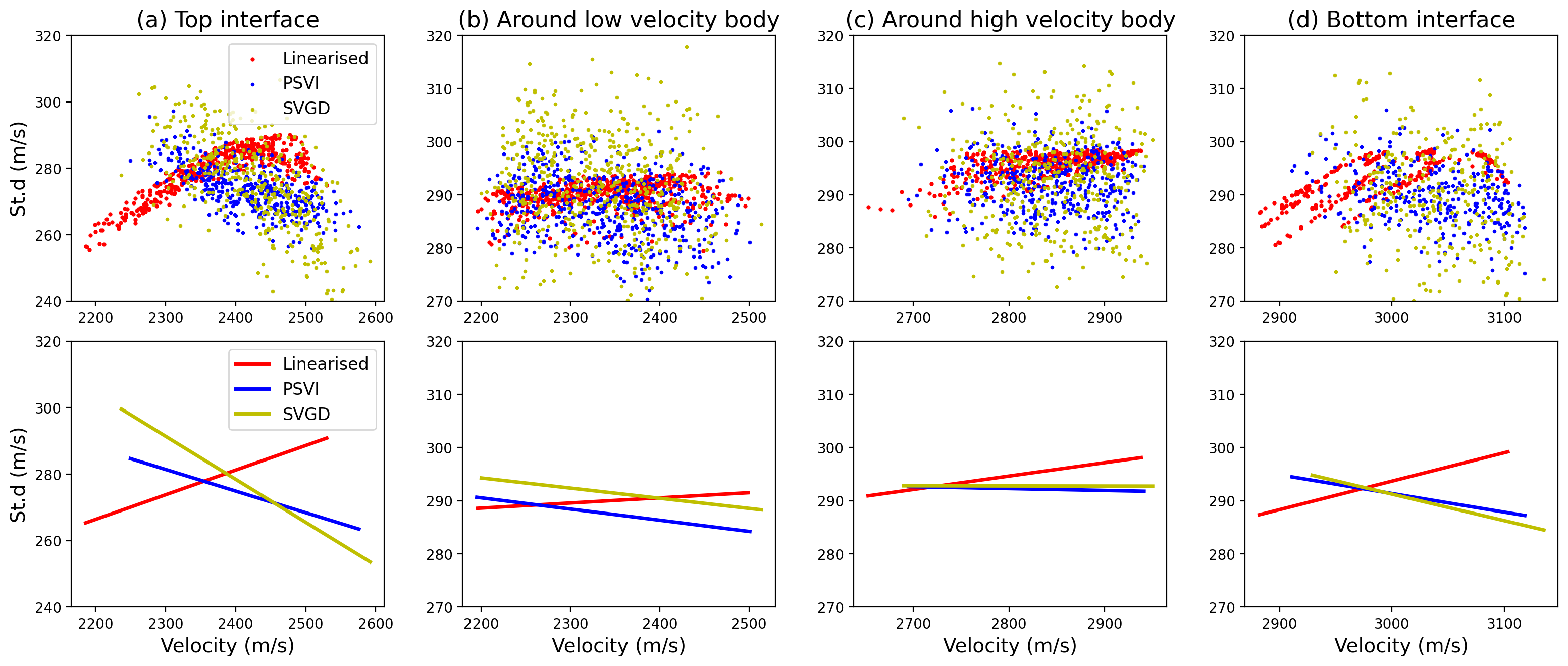}
	\caption{Scatter plots (top row) and the corresponding linear trends (bottom row) of inversion results displayed in Figure \ref{fig:layered_mean_std_10hz_s2r10}. Key as in Figure \ref{fig:layered_mean_std_scatter}.}
	\label{fig:layered_mean_std_scatter_s2r10}
\end{figure}

Finally, we perform FWI using far less informative data with a dominant frequency of 10 Hz but collected using a sparse acquisition geometry that contains only 2 sources and 10 receivers. Figures \ref{fig:layered_mean_std_10hz_s2r10} and \ref{fig:layered_mean_std_scatter_s2r10} show inversion results obtained using this waveform dataset. In each panel in the top row in Figure \ref{fig:layered_mean_std_10hz_s2r10}, red stars and white triangles denote the source and receiver locations, respectively. The three mean velocity maps are similar and closely resemble the prior mean velocity model displayed in Figure \ref{fig:layered_true_initial}b. We observe some reduction in standard deviation values around the source and receiver locations, the top layer interface, and within the low velocity anomaly. Particularly in Figure \ref{fig:layered_mean_std_scatter_s2r10}a, opposite trends of how posterior standard deviation values vary with their mean velocity values around the top interface, occur using the linearised and two nonlinear methods. These features are mainly because data have relatively higher sensitivity at these locations compared to elsewhere. \citet{zhao2025uncertainty} demonstrated that strong prior information about parameter uncertainties is essential in order to refine uncertainty estimates if data are collected using sparse seismic surveys. Nevertheless, the three standard deviation maps are generally consistent, and in Figures \ref{fig:layered_mean_std_scatter_s2r10}b, c and d, their corresponding values are all close to the prior uncertainties (300 m/s), meaning that the posterior pdf's are dominated by the prior distribution, given the limited data input. This example shows that both linearised and nonlinear methods tend to provide similar uncertainty results if data are non-informative.

\subsection{Modified Marmousi Model}
\begin{figure}
	\centering\includegraphics[width=\textwidth]{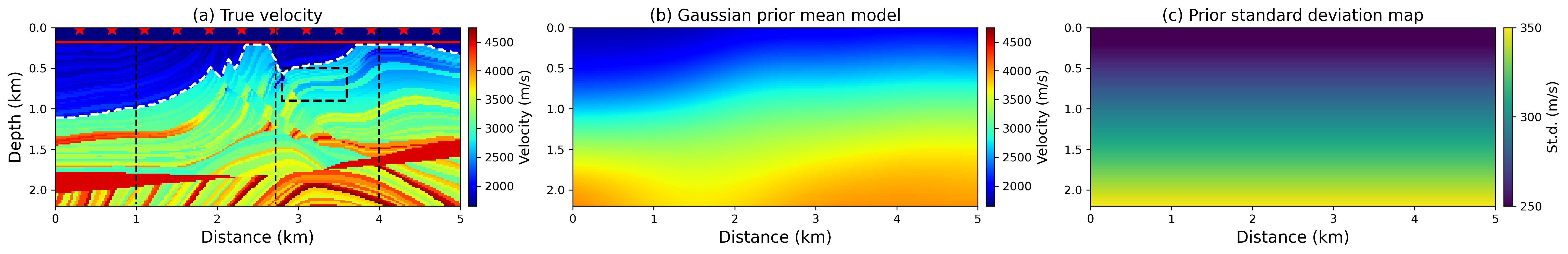}
	\caption{(a) P-wave velocity of the modified Marmousi model used in the fourth FWI example. Source locations are indicated by red stars and a receiver line is marked by a red line. Dashed white curve denotes an interface with high velocity contrast. Vertical dashed black lines display locations of three profiles used to compare the inversion results in later figures. A dashed black box shows a subregion used to interrogate uncertainty results in Section \ref{sec:interrogation}.
		(b) and (c) Gaussian prior mean and standard deviation values used in this example. Mean model in (b) is obtained by smoothing the true velocity model in (a), and standard deviation values in (c) increase linearly with depth from 250 m/s to 350 m/s.}
	\label{fig:marmousi_true_initial}
\end{figure}

In the fourth example, we perform a more realistic FWI problem. As shown in Figure \ref{fig:marmousi_true_initial}a, the true velocity model is modified from the original Marmousi model \cite{martin2006marmousi2} by reducing velocity values above a dashed white line to create an interface with high impedance contrast (also high velocity contrast since we assume a constant density value in this example). Hereafter, we refer to this interface as the main interface to keep the text concise. True velocity profiles along three vertical dashed black lines in Figure \ref{fig:marmousi_true_initial}a are shown by solid green lines in the top row in Figure \ref{fig:marmousi_traces}, illustrating the large velocity contrast. 

\begin{figure}
	\centering\includegraphics[width=\textwidth]{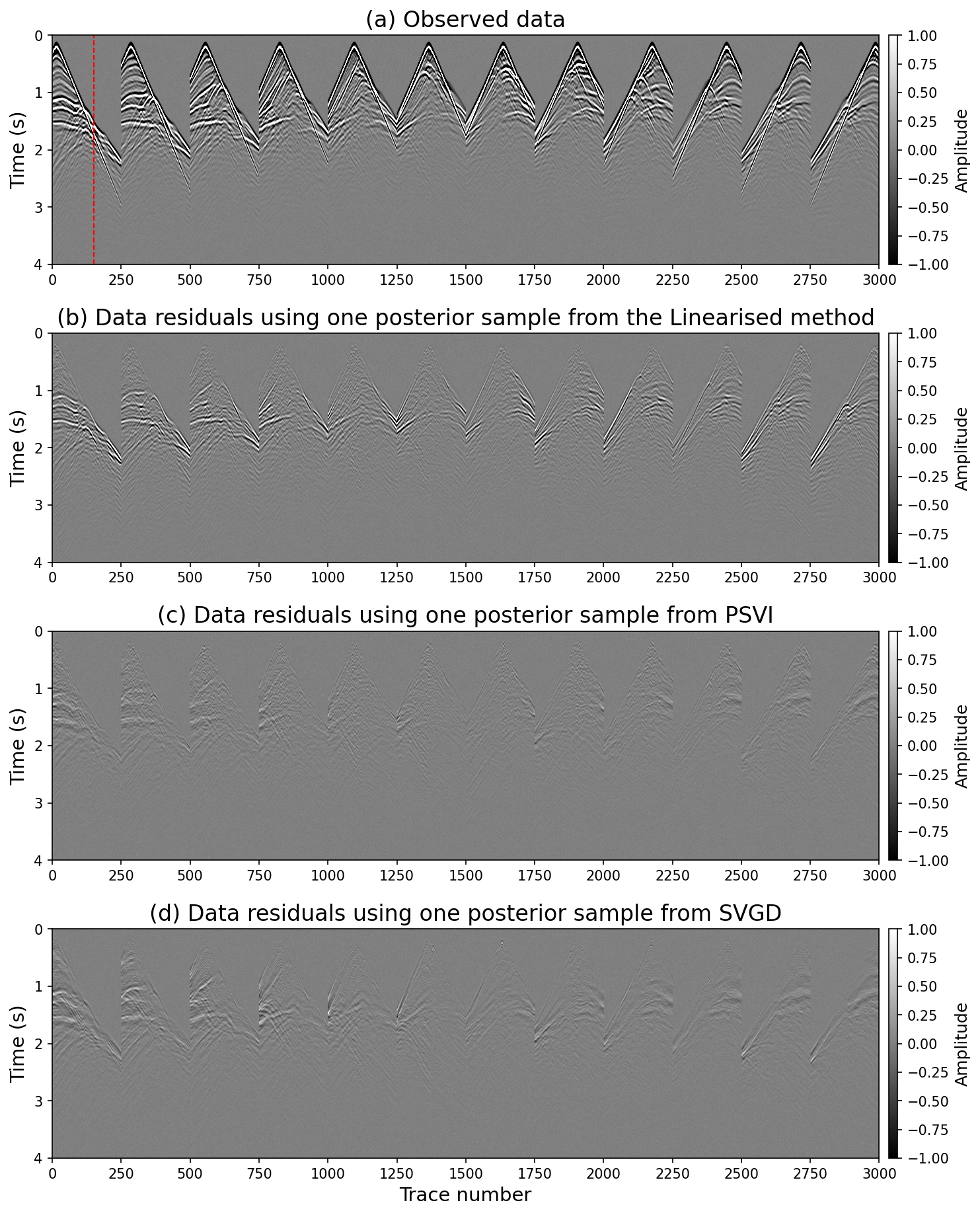}
	\caption{(a) Observed (noisy) waveform data used in the Marmousi example. A vertical dashed red line highlights one seismogram compared in Figure \ref{fig:marmousi_data_trace625_comparison} and in Appendix \ref{ap:datafit}. (b), (c) and (d) Residuals between observed data in (a) and synthetic waveform data simulated by one posterior sample from linearised, PSVI and SVGD methods. Same color scale is applied in each panel.}
	\label{fig:marmousi_observed_data_residual}
\end{figure}

The true model is discretised into 110 $\times$ 250 regular grid cells with a cell size of 20 m in both directions. Twelve sources (red stars in Figure \ref{fig:marmousi_true_initial}a) are placed on the surface, and a receiver line with 250 receivers (horizontal red line) is placed on the seabed at 200 m depth. We employ a Ricker wavelet with 10 Hz dominant frequency to generate a waveform dataset from the true velocity model, and add Gaussian random noise with a standard deviation value equivalent to 1\% of the average of the maximum values of all traces. The resultant data are used as the observed waveform data in this example, as shown in Figure \ref{fig:marmousi_observed_data_residual}a.

Similarly to the previous example, we smooth the true velocity model to obtain a Gaussian prior mean velocity model as shown in Figure \ref{fig:marmousi_true_initial}b. To reflect a realistic situation in which a good prior model is not available, in this example we use a large smoothing window. The prior mean model has almost none of the details in the true model. In Figure \ref{fig:marmousi_true_initial}c, a horizontally constant Gaussian prior uncertainty model is defined, with uncertainty values increasing linearly from 250 m/s at the top to 350 m/s at the bottom of the model. Velocity values above the receiver line (red line in Figure \ref{fig:marmousi_true_initial}a) are fixed during inversion. A diagonal data covariance matrix $\bold\Sigma_{\mathbf{d}}$ is defined for the likelihood function, with diagonal values equal to the magnitude of random noise added to the noise-free waveform data -- a situation in which data noise level is estimated accurately.

\subsubsection{\textbf{Inversion Results}}
\begin{figure}
	\centering\includegraphics[width=\textwidth]{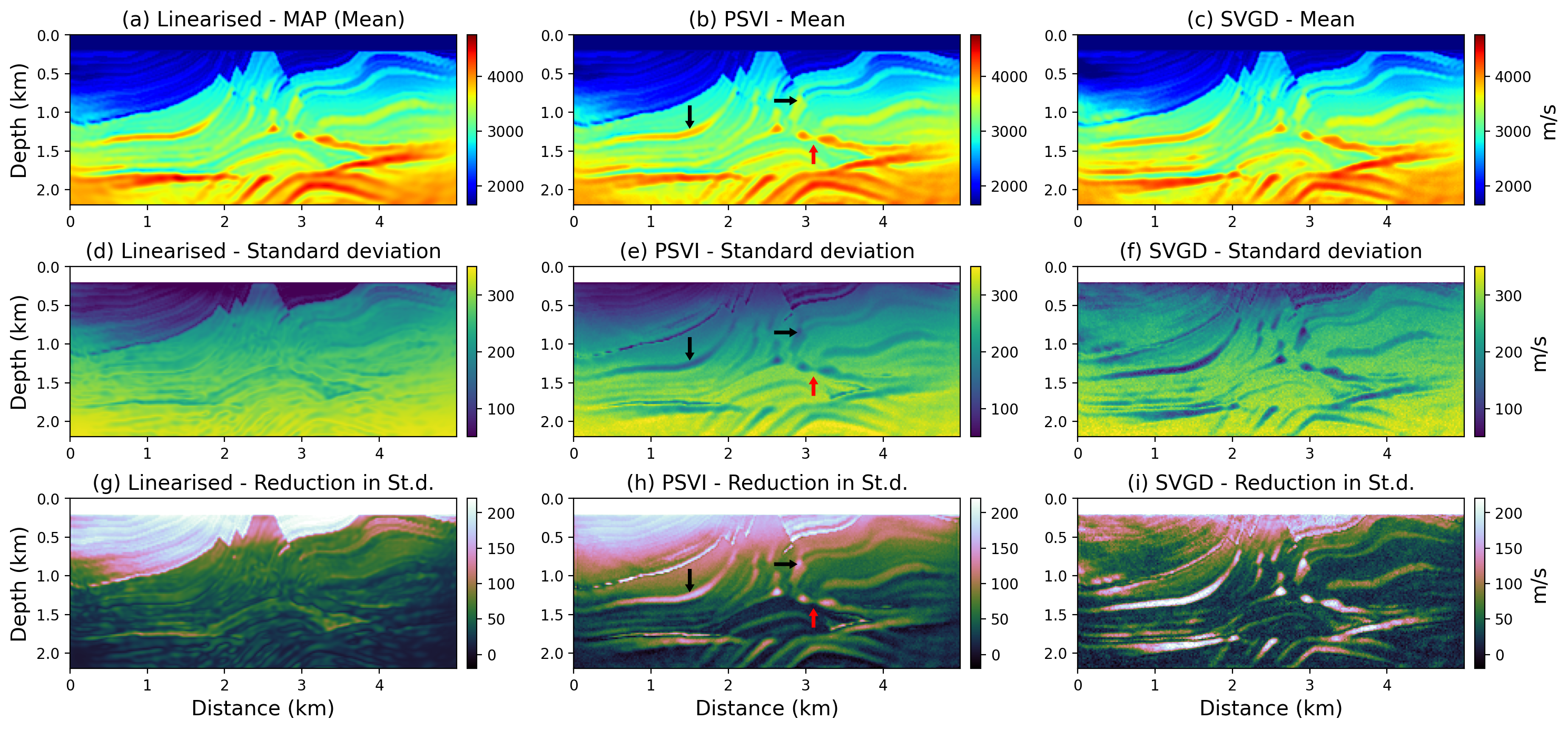}
	\caption{Posterior mean model (top row), standard deviation map (middle row) and reduction in standard deviation values (bottom row) obtained using linearised method (left column), PSVI (middle column) and SVGD (right column). In the left column, panel (a) shows the MAP solution from the linearised inversion, which is equivalent to the Gaussian posterior mean model. In the bottom row, the reduction in standard deviation values is calculated by subtracting the posterior standard deviation values in the middle row from the prior standard deviation in Figure \ref{fig:marmousi_true_initial}c at each point. Since velocity values in the top 200 m are fixed at their true values during inversion, the corresponding regions are left blank in the middle and bottom rows. Panels in the centre and right columns are obtained similarly.}
	\label{fig:marmousi_mean_std_10hz}
\end{figure}

We apply linearised, PSVI and SVGD methods to this FWI problem to estimate solution uncertainties. For the linearised inversion, we first invert 4 Hz waveform data, and its result is used as the reference model for the inversion of 10 Hz data. Figure \ref{fig:marmousi_mean_std_10hz}a presents the final MAP solution obtained. We use 20,000 random vectors to sample the Hessian matrix, and use the randomised singular value decomposition algorithm to estimate the Gaussian posterior covariance matrix. The posterior standard deviation map is displayed in Figure \ref{fig:marmousi_mean_std_10hz}d. For both PSVI and SVGD, we invert the 10 Hz waveform data directly. The posterior mean and standard deviation maps are displayed in Figures \ref{fig:marmousi_mean_std_10hz}b \& e for PSVI and \ref{fig:marmousi_mean_std_10hz}c \& f for SVGD, respectively.

\begin{figure}
	\centering\includegraphics[width=0.8\textwidth]{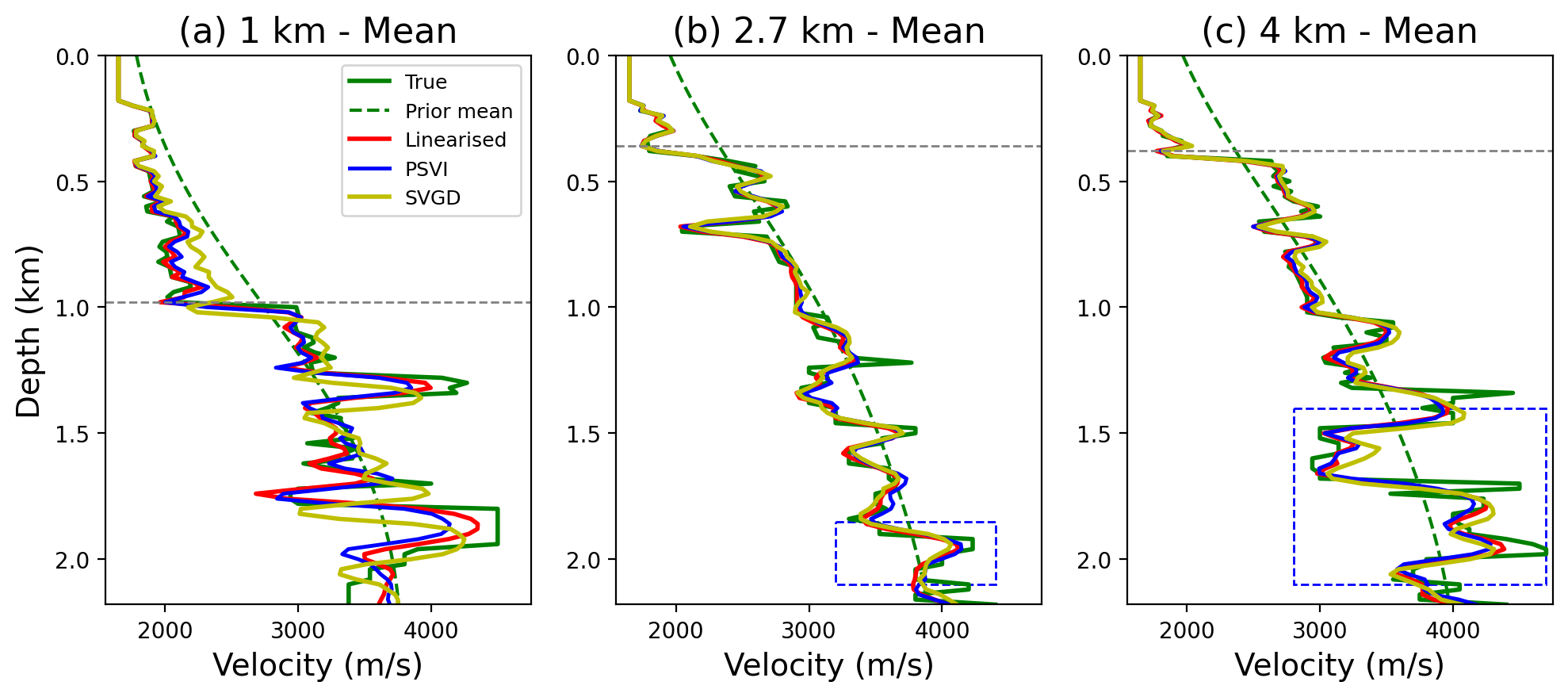}
	\centering\includegraphics[width=0.8\textwidth]{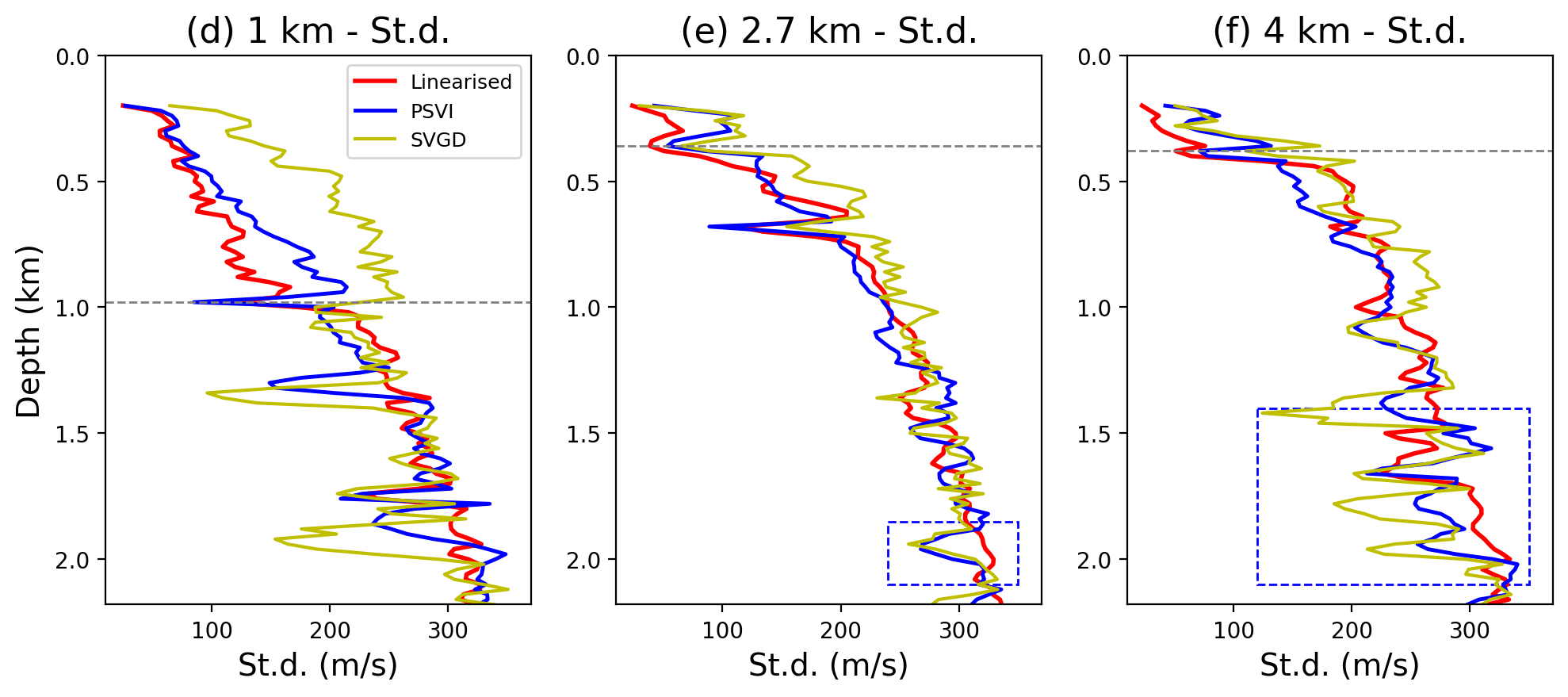}
	\centering\includegraphics[width=0.8\textwidth]{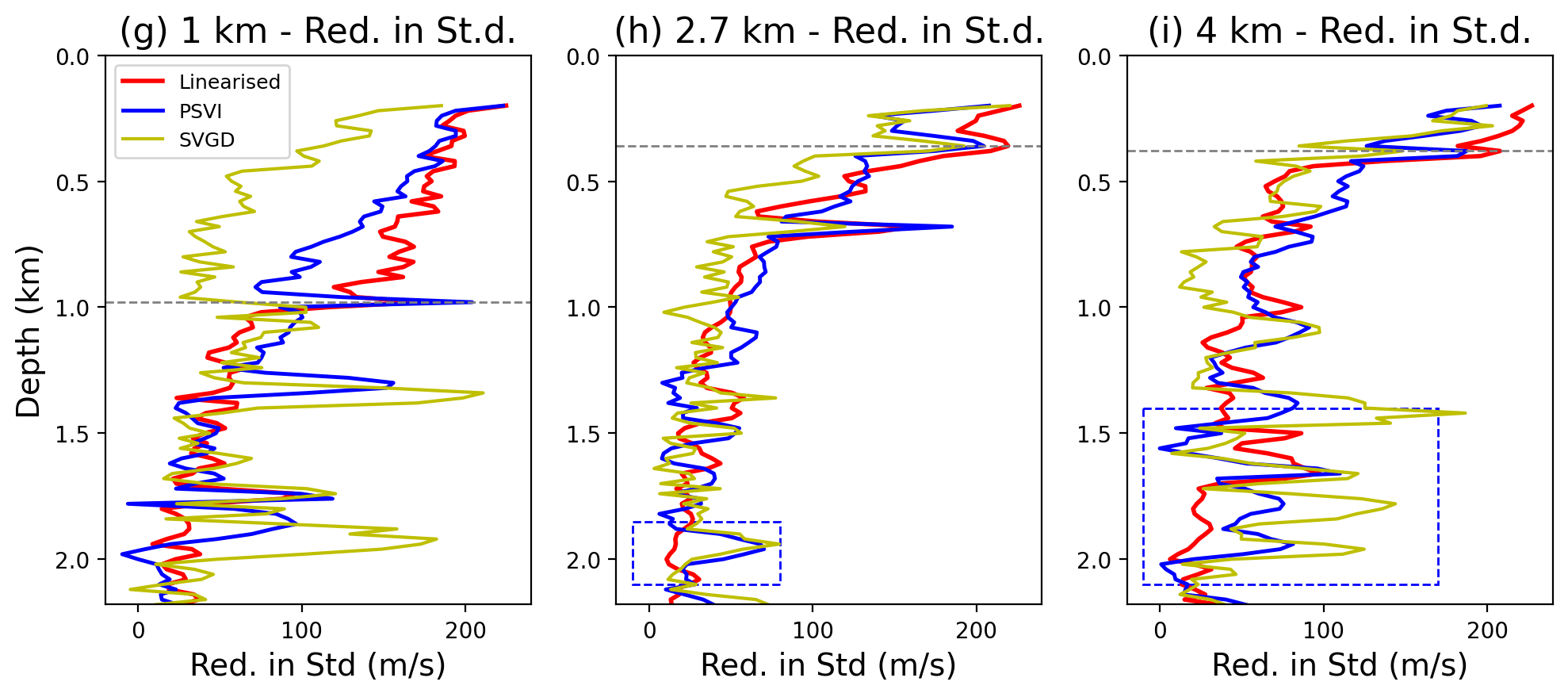}
	\caption{Three vertical profiles of the inversion results extracted at 1 km, 2.7 km and 4 km in each column, respectively. Their locations are represented by vertical dashed black lines in Figure \ref{fig:marmousi_true_initial}a. (a), (b) and (c) Posterior mean velocity profiles, (d), (e) and (f) posterior standard deviation (`St.d.' in the figure) profiles, and (g), (f) and (i) the corresponding profiles showing the reduction in standard deviation (`Red. in St.d.') values between the prior and posterior uncertainties. Horizontal dashed grey lines show the depth of the main interface discussed in the main text, and dashed blue boxes highlight regions within which uncertainty structures from linearised and nonlinear methods show opposite polarities of value variations.}
	\label{fig:marmousi_traces}
\end{figure}

The linearised MAP solution and the posterior mean maps from PSVI and SVGD present consistent structures, generally resembling the true velocity map in Figure \ref{fig:marmousi_true_initial}a. To compare these results in more detail, we plot three vertical profiles extracted from the mean (MAP) models at distances of 1 km, 2.7 km and 4 km, in Figures \ref{fig:marmousi_traces}a, b and c. Compared to the prior mean profiles (dashed green lines), the three sets of posterior mean traces provide more detail and show good consistency with the true velocity profile (green line). However in Figure \ref{fig:marmousi_traces}a, the results from SVGD show significant errors (larger deviations from the true model) compared to those from the other two methods. In Figure \ref{fig:marmousi_mean_std_10hz}c, subtle differences can be observed on the left part of the model at horizontal distances between 0--2 km, compared to Figures \ref{fig:marmousi_mean_std_10hz}a and b. This is possibly because SVGD fails to recover the correct low-wavenumber components of the true model -- a phenomenon referred to as \textit{cycle skipping} in FWI. In SVGD, 600 model samples (often referred to as \textit{particles} in this method) are used to approximate the statistics of the posterior distribution with a dimensionality of 25,000, which are likely to be the cause, since this number may be insufficient to capture even the main components of variation in the solution.

In Figure \ref{fig:marmousi_mean_std_10hz}d, we observe low uncertainty values above the main interface (denoted by the dashed white curve in Figure \ref{fig:marmousi_true_initial}a), and higher uncertainty values with few spatial variation below the interface. Note that the linearised method estimates local posterior uncertainties surrounding the MAP solution (Figure \ref{fig:marmousi_mean_std_10hz}a) by assuming linearised wave physics. In FWI, linearising wave physics involves using the born approximation, which states that given a background reference Earth model (the MAP solution in this case), only the first-order scattering, rather than full scattering, is used to predict seismic wavefields generated by a source and recorded by receivers on the surface. In the MAP solution, the main interface is recovered with reasonable accuracy, meaning that for this background model a large portion of the incident wave energy generated by surface seismic sources is reflected back at this interface, with relatively limited wave energy refracted to the structures below. Therefore, the scattered wavefield energy (the perturbation of waveform data compared to synthetic data predicted by the MAP solution) created by a scatterer (a velocity perturbation) that is above the main interface would be larger than that created by a scatter below the interface. In other words, given constant data uncertainties, the extent to which velocity values can vary above this interface should be smaller than for those below the interface, under a linearised wave physics assumption. Below the interface in Figure \ref{fig:marmousi_mean_std_10hz}d, we therefore observe a significant increase in posterior uncertainties over the entire model. Note that these uncertainties are not obtained from the observed waveform data through the correct, nonlinear forward function; as we will show below, these uncertainties are rather inaccurate.

The posterior standard deviation maps obtained from the two nonlinear methods (Figures \ref{fig:marmousi_mean_std_10hz}e and f) show significantly different structures compared to that from the linearised method. These maps contain more detailed and more clearly defined geometrical uncertainty structures that more closely resemble the true shape of the velocity variations below the main interface. The posterior uncertainty values above and below the main interface do not change much, and they change significantly only at locations where the corresponding velocity values vary (significantly). Particularly, lower uncertainty structures at deeper part of the model below the main interface are more obvious compared to the linearised inversion results. Figure \ref{fig:marmousi_mean_std_10hz} thus demonstrates that nonlinear wave physics (i.e., the full scattering effect) enables regions of high and low uncertainty to be constrained more effectively than does the linearised physics (first-order scattering). Similarly in electrical resistivity tomography, \citet{galetti2018transdimensional} demonstrated that including fully nonlinear physics allows deeper parts of the Earth to be imaged with the same data.

We also calculate the reduction in standard deviation values, defined as the differences between the prior standard deviation (Figure \ref{fig:marmousi_true_initial}c) and the posterior standard deviation values from the three methods, which are displayed in Figures \ref{fig:marmousi_mean_std_10hz}g, h and i. These maps remove the effect of the increasing trend in prior uncertainties with depth and highlight spatial variations of uncertainty reduction achieved by the introduction of the waveform data. Similarly to the posterior standard deviation maps in the middle row in Figure \ref{fig:marmousi_mean_std_10hz}, significantly different results are obtained from the linearised and nonlinear inversion methods. In Figure \ref{fig:marmousi_mean_std_10hz}g, large reductions are observed above the main interface, but below the interface the reduction is generally low with few spatial variations. 

The results from the two nonlinear methods show consistent patterns of spatial variations, thereby cross-validating each other. In particular, below the main interface where data sensitivity is lower compared to the regions above the interface, posterior uncertainties decrease (and reduction values increase) around multiple layer boundaries where velocity values increase. These are highlighted by black arrows in Figure \ref{fig:marmousi_mean_std_10hz}. Conversely, uncertainties increase (reduction values decrease) in area where velocities decrease, as indicated by red arrows in Figure \ref{fig:marmousi_mean_std_10hz}. These features align with those illustrated in the previous layered example (using 10 Hz data in Figure \ref{fig:layered_mean_std}), particularly around the boundaries of the low and high velocity anomalies. 

To further support the above statements, in the second and third rows in Figure \ref{fig:marmousi_traces} we plot the profiles of the posterior standard deviation and the corresponding reduction values along the same vertical profiles as in the first row (mean velocity values). Generally, uncertainty values increase (values of reduction in uncertainties decrease) with depth, as sensitivity of waveform data diminishes at greater depths. Horizontal dashed grey lines show the locations of the main interface, above which uncertainty variations from the three methods are similar (albeit with varying amplitudes) since the velocity values and their uncertainties are more tightly constrained by the waveform data. For example, around the interface (dashed grey lines), low uncertainties are observed across all three sets of results. 

Below the interface, spatial variations of the uncertainty (and reduction) values are smaller in the results from the linearised method than those from the two nonlinear methods, due to the linearisation of wave physics as discussed above. Around several layer interfaces, we observe that uncertainty values from the two nonlinear methods increase (thus reduction in uncertainties decreases) with decreasing velocity values, consistent with those displayed in the layered model example (Figure \ref{fig:layered_mean_std_scatter}). These uncertainty variations show opposite trends in the linearised results, where clear spatial variations are observable. Two such examples are highlighted within the dashed blue boxes in the right two columns in Figure \ref{fig:marmousi_traces}.

To conclude, this example shows that in realistic FWI problems in which subsurface structures are complex, data are contaminated by noise and accurate prior information is not available, uncertainty estimates from linearised and nonlinear methods can be significantly different, especially in situations in which interfaces with high impedance contrast exist.

\subsubsection{\textbf{Uncertainty Appraisal and Interrogation}}
\label{sec:interrogation}

Given the above significantly different uncertainty results, we test their accuracy by examining whether they describe various Earth models that fit observed waveform data, and whether they provide unbiased information that can be used to interrogate subsurface structures accurately. 

We first assess data fit by comparing observed waveform data with data simulated from posterior samples. We draw five random samples from each of the three posterior pdf's, and perform forward simulations with these samples. Figures \ref{fig:marmousi_observed_data_residual}b, c and d display data residuals (differences) between the observed data (Figure \ref{fig:marmousi_observed_data_residual}a) and synthetic data simulated by one posterior sample from the linearised method, PSVI and SVGD, respectively. Significant data misfits are observed from the linearised method, meaning that this posterior sample can not generate synthetic data that fit the observed data well. Data residuals calculated from samples from PSVI and SVGD are similar and are significantly smaller than those from the linearised method. To prove that this remains true across different posterior samples, in Appendix \ref{ap:datafit} we display data residuals calculated from different samples drawn from the three posterior pdf's.

\begin{figure}
	\centering\includegraphics[width=\textwidth]{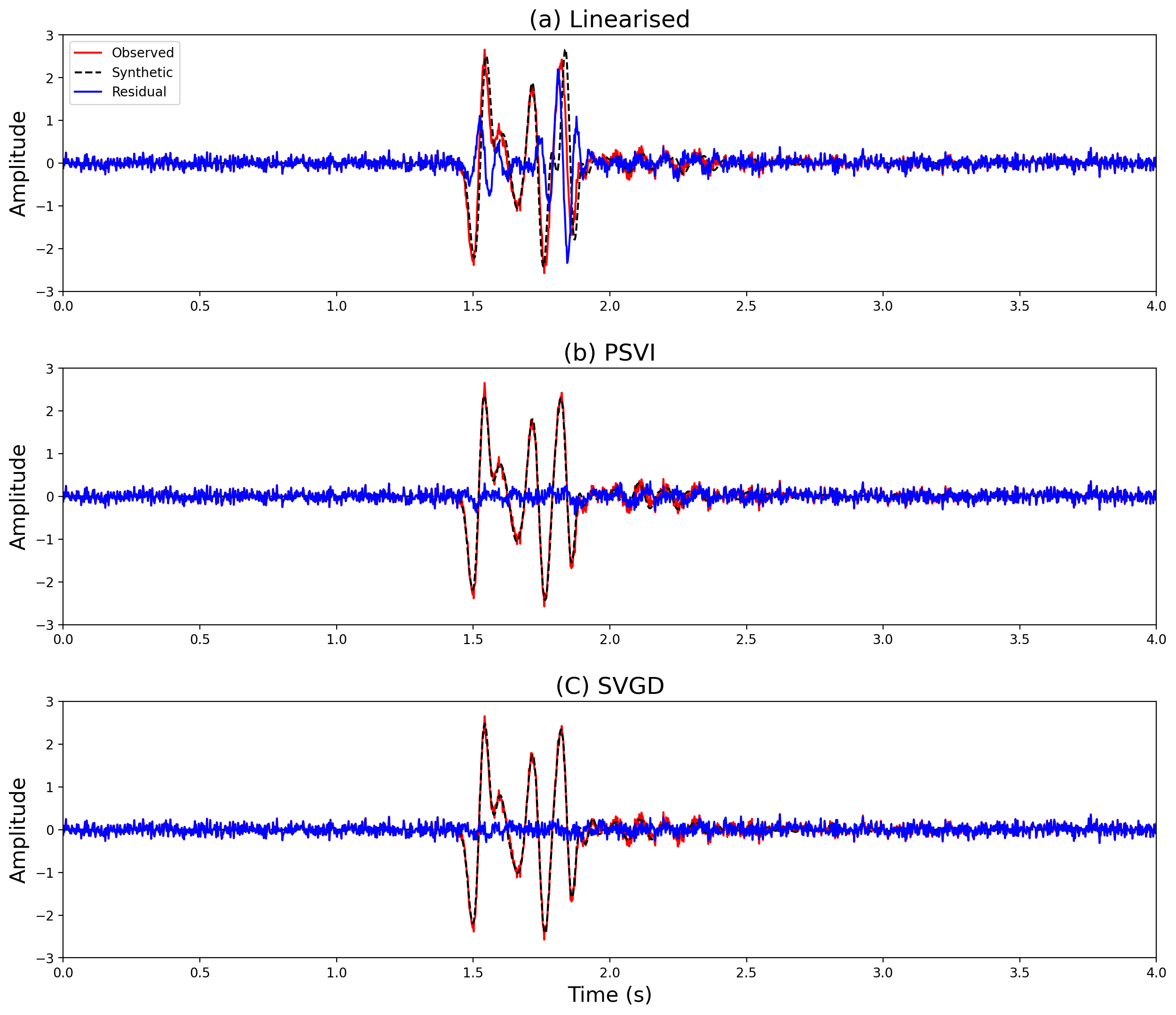}
	\caption{Comparison of one observed seismogram (red curve) and the corresponding synthetic seismogram (dashed black curve) simulated by one posterior sample from (a) linearised method, (b) PSVI and (c) SVGD. The seismogram is highlighted by a dashed red line in Figure \ref{fig:marmousi_observed_data_residual}a. In each panel, a blue curve shows the differences (data residuals) between the observed and synthetic data. Further examples from different posterior samples are given in Appendix \ref{ap:datafit}.}
	\label{fig:marmousi_data_trace625_comparison}
\end{figure}

\begin{figure}
	\centering\includegraphics[width=\textwidth]{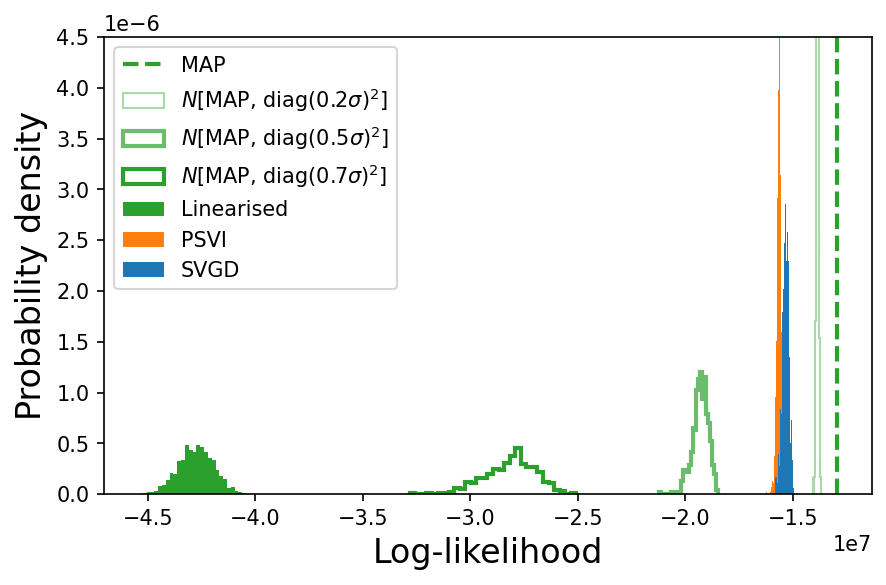}
	\caption{Solid histograms with different colours: posterior histograms of the logarithmic likelihood values of 500 posterior samples obtained from different inversion results - higher value reflect better data fits. Hollow green histograms with different weights: histograms of the logarithmic likelihood values of 500 samples from Gaussian distributions $\mathcal{N}(\text{MAP}, diag(\alpha\sigma)^2)$, centred on the linearised MAP solution with different covariance matrices $diag(\alpha\sigma)^2$ with diagonal values $(\alpha\sigma)^2$, where $\alpha = \{0.2, 0.5, 0.7\}$ and $\sigma$ are the linearised posterior standard deviation values displayed in Figure \ref{fig:marmousi_mean_std_10hz}d. Dashed vertical green line denotes the logarithmic likelihood value of the MAP solution. The logarithmic likelihood value of each sample is calculated by equation \ref{eq:likelihood}.}
	\label{fig:marmousi_loglike}
\end{figure}

For a clearer comparison, in Figure \ref{fig:marmousi_data_trace625_comparison} we plot one observed seismogram (highlighted by the dashed red line in Figure \ref{fig:marmousi_observed_data_residual}a), alongside the simulated seismograms and the corresponding data residuals from the three methods. Obvious phase differences are observed between the observed (red line) and synthetic (dashed black line) seismograms in Figure \ref{fig:marmousi_data_trace625_comparison}a. Data residuals (blue line) are thus significant. On the other hand, synthetic seismograms in Figures \ref{fig:marmousi_data_trace625_comparison}b and c generated from PSVI and SVGD fit the observed data more accurately. Data residuals from these two nonlinear methods are nearly at the same amplitude as that of background data noise, indicating a good data fit. In Appendix \ref{ap:datafit}, we compare the same observed seismogram with simulated ones, and show their residuals from different posterior samples. These results prove that data fits from the two nonlinear methods are generally better than those from the linearised method.

In addition, we draw 500 samples from the three posterior distributions, simulate their synthetic waveform data, and calculate the logarithmic likelihood value for each posterior sample using equation \ref{eq:likelihood}. In Figure \ref{fig:marmousi_loglike}, solid histograms with different colours show distributions of the calculated logarithmic likelihood values. Those from samples obtained using the linearised method are significantly lower than those from the two nonlinear methods, demonstrating that the linearised method can not provide accurate uncertainty results that fit observed data. The vertical dashed green line in Figure \ref{fig:marmousi_loglike} denotes the logarithmic likelihood value calculated from the MAP solution. 

However, the linearised (Gaussian) posterior distribution is centred on the MAP solution which (if we find the correct MAP solution) should provide the highest likelihood value. There should therefore exist some neighbourhood of the MAP solution in which samples fit the data similarly well. We generate three sets of 500 samples in the MAP's neighbourhood by sampling three Gaussian distributions $\mathcal{N}(\text{MAP}, diag(\alpha\sigma)^2)$ each with a mean vector of the MAP solution and a diagonal covariance matrix $diag(\alpha\sigma)^2$. The diagonal values are $(\alpha\sigma)^2$, where $\alpha = \{0.2, 0.5, 0.7\}$ and $\sigma$ are the linearised posterior standard deviation values displayed in Figure \ref{fig:marmousi_mean_std_10hz}d. Three hollow green histograms with different levels of transparency in Figure \ref{fig:marmousi_loglike} show the resulting logarithmic likelihood values. These increase as $\alpha$ decreases since the corresponding samples are closer to the MAP solution within parameter space; the values from $\mathcal{N}(\text{MAP}, diag(0.2\sigma)^2)$ are higher than those from PSVI and SVGD to the point where all of three sets of the likelihood values are higher than those represented by the green solid histogram. However, samples randomly drawn from the full linearised posterior distribution consistently show low likelihood values (solid green histogram). This is because samples of a high-dimensional Gaussian distribution almost never lie close to its mean model, since the total probability mass of samples that are close to the Gaussian mean vector is negligible compared to the probability mass of samples that are \textit{not} close to the mean \cite{curtis2001prior}. This implies that it is important that any Gaussian approximation is correlated in directions that represent as accurately as possible those in the true, global posterior distribution. Figure \ref{fig:marmousi_loglike} shows that PSVI (a log-transformed Gaussian) does so far more accurately than linearised uncertainty analysis, and in this case provides data fits that are similar to those from SVGD.

Next, we use the obtained uncertainty information to interrogate subsurface structures using \textit{interrogation theory} \cite{arnold2018interrogation}. Interrogation theory is designed to answer low-dimensional, high-level scientific questions about the Earth in an unbiased manner. Examples of the types of questions asked might be: \textit{What is the total volume of CO$_2$ stored in a subsurface reservoir? What is the probability of an earthquake occurring in a specific region?} In interrogation theory, the optimal (minimally biased) answer $a^*$ to a specific question $Q$ is estimated by calculating the following expectation term:
\begin{equation}
	a^* = \mathbb{E}_{p(\mathbf{m}|\mathbf{d}_{obs})}[T(\mathbf{m}|Q)] = \int_\mathbf{m} T(\mathbf{m}|Q)p(\mathbf{m}|\mathbf{d}_{obs})\ d\mathbf{m},
	\label{eq:optimal_answer}
\end{equation}
The expectation is taken with respect to the posterior distribution $p(\mathbf{m}|\mathbf{d}_{obs})$ of model parameters $\mathbf{m}$. Therefore, the final optimal answer incorporates uncertainty information from the inversion results. Term $T(\mathbf{m}|Q)$ is a target function conditioned on the question $Q$ of interest. It is defined to map a high dimensional model vector $\mathbf{m}$ into a low dimensional target function value $t$ in a target space $\mathbb{T}$, within which the question $Q$ can be answered directly. In such cases, the optimal answer in equation \ref{eq:optimal_answer} is simply the expectation of the posterior target function.

\begin{figure}
	\centering\includegraphics[width=\textwidth]{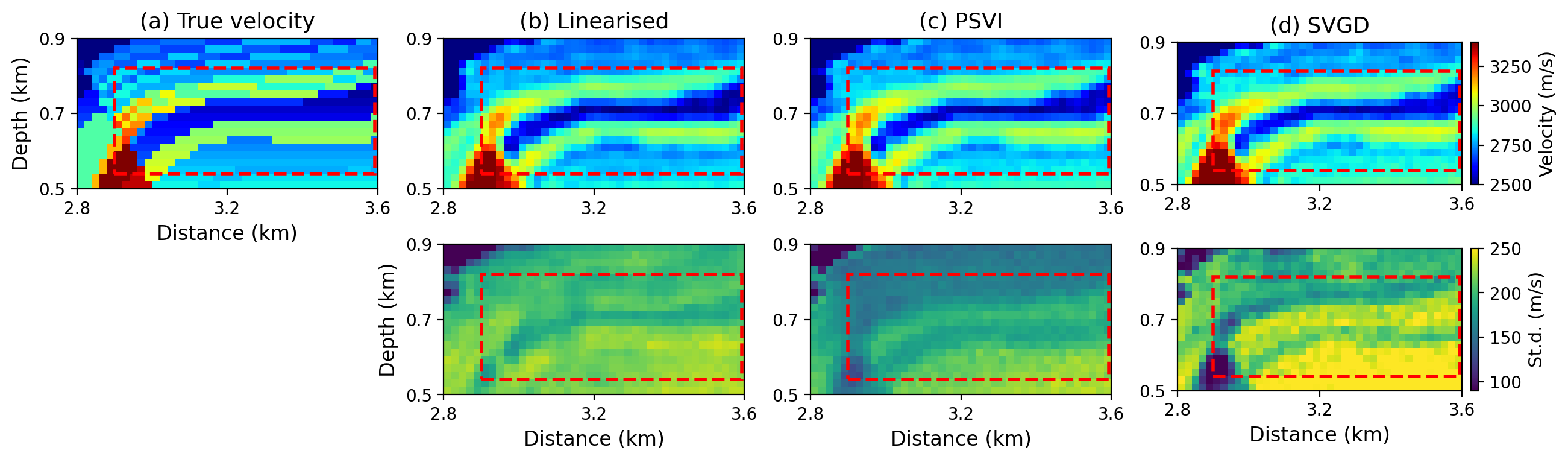}
	\caption{(a) True velocity model within a dashed black box in Figure \ref{fig:marmousi_true_initial}a, and the corresponding posterior mean (top row) and standard deviation (bottom row) maps of the inversion results obtained using (b) the linearised method, (c) PSVI and (d) SVGD. In each panel, a dashed red box marks a low velocity body, and its size is estimated by applying interrogation theory.}
	\label{fig:marmousi2_small_for_interrogation}
\end{figure}

We focus on a subregion of the velocity model highlighted by a dashed black box in Figure \ref{fig:marmousi_true_initial}a. The corresponding true velocity structure is shown in Figure \ref{fig:marmousi2_small_for_interrogation}a. Figures \ref{fig:marmousi2_small_for_interrogation}b, c and d display the corresponding posterior mean (top row) and standard deviation (bottom row) maps obtained from the three methods (magnified versions of results above). In Figure \ref{fig:marmousi2_small_for_interrogation}a, a low velocity body is observed inside of a dashed red box, and it is recovered reasonably accurate in the three posterior mean velocity maps. However, posterior uncertainty structures differ significantly between the linearised and two nonlinear methods. Specifically, results from both PSVI and SVGD show high uncertainty values within the low velocity body and low uncertainty values in the surrounding high velocity regions. Completely different uncertainty structures are obtained from the linearised method: low uncertainties within the low velocity body and high uncertainties within the surrounding high velocity regions. These different results could lead to different (and potentially biased) interpretation concerning subsurface properties. 

To demonstrate, we use the three sets of results to interrogate subsurface information by answering the following question: \textit{what is the size of the low velocity body within the dashed red box?} Previously, volume-related questions were answered by applying interrogation theory to seismic imaging results from travel time tomography \cite{zhao2022interrogating} and FWI \cite{zhang2021interrogation, zhao2024bayesian}. Similarly, we define a target function $T(\mathbf{m})$ to represent the size (area, since this is a spatially 2D example) of the largest continuous low velocity body within the considered region. This function transforms a high dimensional velocity model $\mathbf{m}$ into a scalar value (size of the low velocity body). 

Note that the evaluation of the target function requires a velocity threshold to distinguish between low and relatively higher velocities. We use the same data-driven method introduced in \citet{zhao2022interrogating} to calculate the least-biased threshold. We first pick some cells that are highly likely to be inside of the low velocity body, and others that are almost definitely outside. The least-biased threshold value satisfies that the expected probability of interior cells being below that value equals the expected probability of exterior cells being above that value. Once we have found this value from the posterior probability distributions, we are able to calculate the target function value for each posterior sample. More details about the interrogation procedure can be found in \citet{arnold2018interrogation} and in the above studies, in particular in \citet{zhao2024bayesian}.

\begin{figure}
	\centering\includegraphics[width=\textwidth]{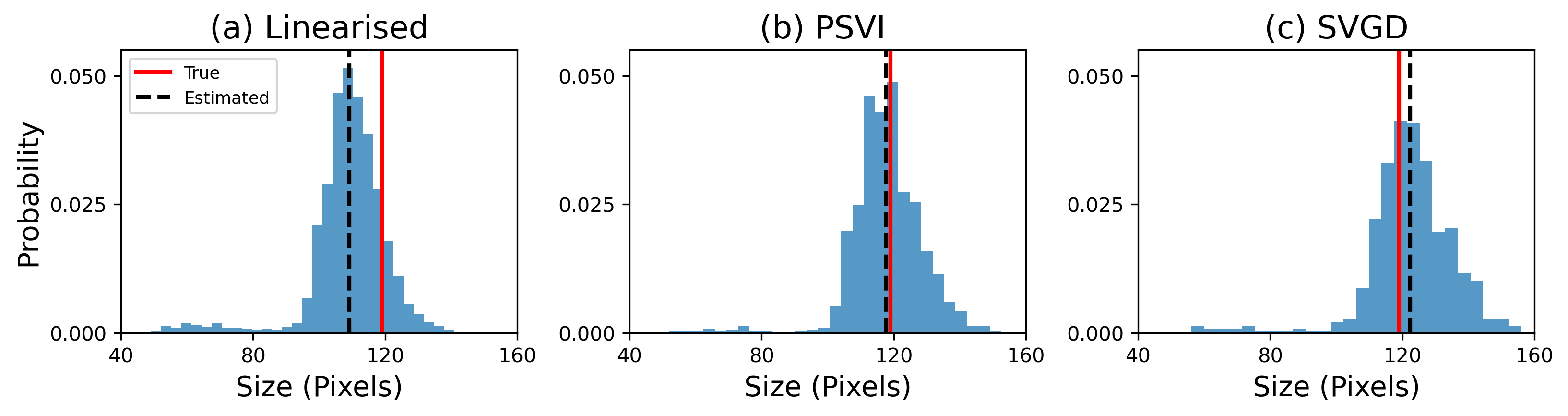}
	\caption{Posterior probability densities of the low velocity body size estimated using inversion results from (a) the linearised method, (b) PSVI, (c) sSVGD, respectively. Red lines denote the true size, and black dashed lines denote the optimal size obtained using interrogation theory.}
	\label{fig:marmousi2_small_interrogation_results}
\end{figure}

Figure \ref{fig:marmousi2_small_interrogation_results} shows the probability densities of the target function values -- i.e., the estimated size of the low velocity body -- derived from different inversion results. The true size is known precisely in this synthetic test, and is marked by a red line in each panel in Figure \ref{fig:marmousi2_small_interrogation_results}. For each histogram, the optimal unbiased answer is calculated using equation \ref{eq:optimal_answer} (i.e., the mean of each histogram), and is indicated by a dashed black line. The interrogation results (both the optimal answers and the overall histograms) from PSVI and SVGD are closer to the true value compared to those obtained from the linearised method; indeed the true value lies within the same histogram bin as the MAP values for both of the nonlinear methods but has an estimated probability of less than half of the MAP value when using linearised methods.

From both the data fit and the interrogation results, we therefore conclude that posterior uncertainties from nonlinear inversion methods are significantly more accurate than those from the linearised method, since the former predicts synthetic waveform data that better fit observed data with smaller misfits and provides more accurate information about subsurface properties.

\section{Discussion}
We have demonstrated that the Gaussian-based linearised method and PSVI yield rather different posterior uncertainty results (with the latter being more accurate), yet both methods employ a Gaussian kernel to approximate the unknown posterior distribution. PSVI is a nonlinear variational inference algorithm that finds an optimal logarithmically-transformed (equation \ref{eq:log_transform}) Gaussian distribution -- the one that minimises the KL-divergence between the posterior pdf and itself -- within the full parameter space. Its results thus describe statistics of the full uncertainties. On the other hand, the linearised method finds a Gaussian distribution to approximate local posterior uncertainty information by linearising the nonlinear forward function (wave physics). This work thus justifies the superiority of PSVI over the linearised method to solve nonlinear inverse problems using a Gaussian kernel.

In the Marmousi example, we deliberately reduced the true velocity values above the dashed white line in Figure \ref{fig:marmousi_true_initial}a to create an interface with high impedance contrast. This enables us to highlight significantly different uncertainty results obtained from linearised and nonlinear methods. Nevertheless, such impedance contrasts are not uncommon in real Earth structures, especially around subsurface salt bodies and hard seabed interfaces at which the impedance contrast can be much larger than that used in our example \cite[e.g., see][]{lomas20233d}. 

\begin{figure}
	\centering\includegraphics[width=\textwidth]{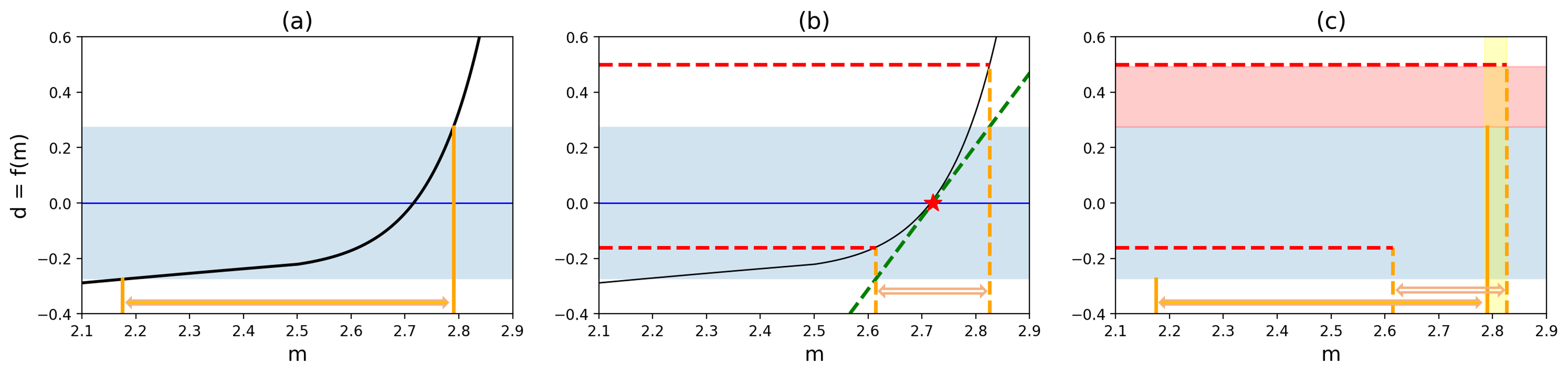}
	\caption{Schematic example comparing linearised and nonlinear uncertainty estimates. In panel (a), a black solid curve shows a nonlinear forward function $d=f(m)$. Horizontal solid blue line denotes an observed datum $d_{obs} = 0$, which has a datum error bar of $\sigma = 0.275$ represented by a horizontal light blue shaded region. Model parameter values between two vertical solid orange lines (solid double-headed arrow) fit the observed datum to within its uncertainty. Panel (b) illustrates the corresponding linearised uncertainty estimates: a red star shows the MAP solution, around which the nonlinear forward function is linearised, represented by a dashed green line. Parameter values between two vertical dashed orange lines (hollow double-headed arrow) denote the linearised uncertainty estimates which clearly deviate from the correct range in panel (a). Datum values between two horizontal dashed red lines denote possible synthetic data predicted by the linearised parameter uncertainties through the nonlinear forward function (solid black curve), which deviate significantly from (and extend beyond) the true data uncertainties in light blue. Panel (c) highlights differences between panels (a) and (b) in both the model and data spaces. Data within a red shaded region do not fit the datum to within observational uncertainties, and are predicted by parameter values highlighted by a yellow shaded region, where the linearised uncertainty estimates extend beyond the true, nonlinear uncertainties.}
	\label{fig:illustration_FWI_nonlinearity_linearised_uncertainty2}
\end{figure}

To further compare uncertainty estimates from linearised and nonlinear methods, we analyse a schematic 1D inversion example in Figure \ref{fig:illustration_FWI_nonlinearity_linearised_uncertainty2}, in which both model parameter and datum are scalar values. In Figure \ref{fig:illustration_FWI_nonlinearity_linearised_uncertainty2}a, a black solid curve shows a nonlinear forward function $d=f(m)$. Assume that we have an observed datum $d_{obs} = 0$ (horizontal solid blue line) with an observational uncertainty error bar of $\sigma = 0.275$ (horizontal light blue shaded area). Model parameter values within two vertical solid orange lines (solid double-headed arrow) fit the observed datum to within its observational uncertainty. These values represent full posterior uncertainties (e.g., estimated by a nonlinear inversion method). Figure \ref{fig:illustration_FWI_nonlinearity_linearised_uncertainty2}b illustrates linearised uncertainty results. We first calculate the MAP solution, as marked by a red star. The forward function is linearised around the MAP solution, resulting in a dashed green line. Given this linearised forward function, we project the observational uncertainty range back to model space and obtain the linearised uncertainty estimates, denoted by parameter values between two vertical dashed orange lines (hollow double-headed arrow). They differ substantially from the nonlinear uncertainties, and are equivalent to differences in linearised versus nonlinear uncertainty estimates in the FWI problems above. We project the linearised parameter uncertainties to datum space through the correct, nonlinear forward function (black curve), giving predicted values between two horizontal dashed red lines. These differ significantly from (and extend outside of) the true datum uncertainty.

To ensure that parameter values from linearised uncertainties fit data to within observational uncertainties, they can not lie within the vertical yellow shaded region in Figure \ref{fig:illustration_FWI_nonlinearity_linearised_uncertainty2}c. Unfortunately, if linearised uncertainty estimates are normally distributed then the hypervolume of parameter space close to the mean (here, the MAP model) is tiny compared to that further afield including the yellow region \cite{curtis2001prior}. This explains why in Figure \ref{fig:marmousi_loglike} almost all random samples from the linearised posterior pdf do not fit the data, while their mean model (MAP solution) has nearly a perfect data fit.

In this work, Gaussian prior distributions were used to simplify the formulation of the linearised posterior distribution \cite{tarantola2005inverse}. It is not clear whether our main conclusions remain applicable when involving more complex and informative prior knowledge. For example, geological prior information modelled by multiple-point geostatistics \cite{mariethoz2014multiple} or deep generative neural networks \cite{mosser2020stochastic, levy2022variational, bloem2022introducing, meles2024bayesian} can be used to regularise FWI problems, potentially simplifying the complexity of the inversion. Future work could investigate the difference between linearised and nonlinear uncertainty estimates when employing informative and sophisticated prior knowledge.

As discussed in Sections \ref{sec:uncertainty_17hz} and \ref{sec:uncertainty_sparse}, posterior uncertainties from both linearised and nonlinear methods can sometimes be similar if data are either highly informative or quite non-informative. While informative data can be collected by optimising survey designs \cite{maurer2010recent}, in practice it is difficult to determine \textit{a priori} whether observed data are sensitive enough to yield similar uncertainty results from linearised and nonlinear methods. In addition, although linearised methods are generally computationally much cheaper than nonlinear methods, they are more sensitive to the starting point of the inversion. The estimated uncertainties are inaccurate and potentially biased, leading to incorrect interpretation of subsurface properties in our example above, and hence potentially to non-robust decision-making. Therefore, practitioners should prioritise nonlinear inversion methods whenever computational resources permit. 

In this study, we only solved spatially 2D constant density, acoustic FWI problems. Multi-parameter FWI such as elastic FWI and spatially 3D FWI problems typically exhibit higher dimensionality and nonlinearity than 2D acoustic FWI. In future, we hope to explore the extent to which uncertainty results from linearised and nonlinear methods differ in these more complex contexts.

\section{Conclusion}
We have performed probabilistic Bayesian FWI and uncertainty quantification using both linearised and fully nonlinear inversion methods, comparing the inversion results obtained from a linearised method with those obtained from two nonlinear variational Bayesian inversion algorithms: PSVI and SVGD. Both linearised and nonlinear methods provide reasonably accurate posterior mean velocity models, but the posterior uncertainties differ significantly. Specifically, around layer interfaces, nonlinear inversion results typically show high uncertainty values on the low velocity side and relatively lower uncertainties on the high velocity side. Such features are absent from linearised uncertainty estimates due to the linearisation of the forward function. By incorporating fully nonlinear wave physics, posterior uncertainty structures extend deeper into the solid Earth using the same data, especially in cases in which the subsurface contains layers with high impedance contrast. In addition, we demonstrate that uncertainties from nonlinear methods are more likely to be accurate than those from the linearised method, since the former provides better data fit and delivers more accurate information about meta-properties of the Earth's interior. While both types of methods yield similar posterior statistics if observed data are either extremely informative or non-informative, it is difficult to know whether a given problem falls into one of these two regimes, given that we never know the true Earth structures. We therefore conclude that nonlinear inversion methods generally provide better uncertainty results and should be preferred for future nonlinear inversion problems.

\section{ACKNOWLEDGMENTS}
We thank the Edinburgh Imaging Project sponsors (BP and TotalEnergies) for supporting this research. For the purpose of open access, we have applied a Creative Commons Attribution (CC BY) licence to any Author Accepted Manuscript version arising from this submission.

\bibliographystyle{plainnat}  % style file is seg.bst
\bibliography{reference}

\appendix

\section{Data Fit from Different Posterior Samples}
\label{ap:datafit}
\begin{figure}
	\centering\includegraphics[width=\textwidth]{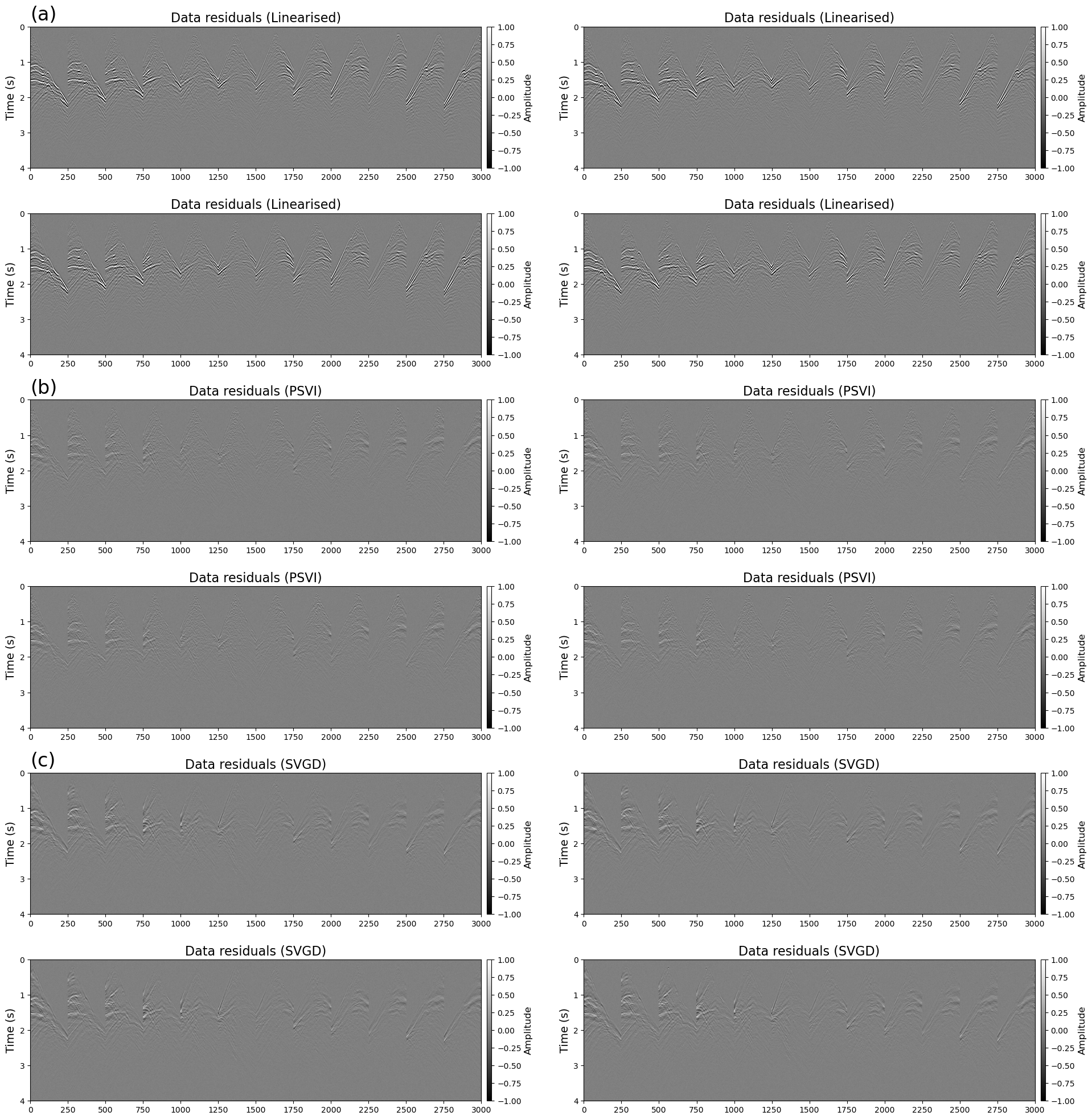}
	\caption{Residuals between observed waveform data in Figure \ref{fig:marmousi_observed_data_residual}a and synthetic data simulated by four additional posterior samples from (a) linearised, (b) PSVI and (c) SVGD inversion results. Key as in Figure \ref{fig:marmousi_observed_data_residual} in the main text.}
	\label{fig:marmousi2_data_residual_from_4samples}
\end{figure}

\begin{figure}
	\centering\includegraphics[width=\textwidth]{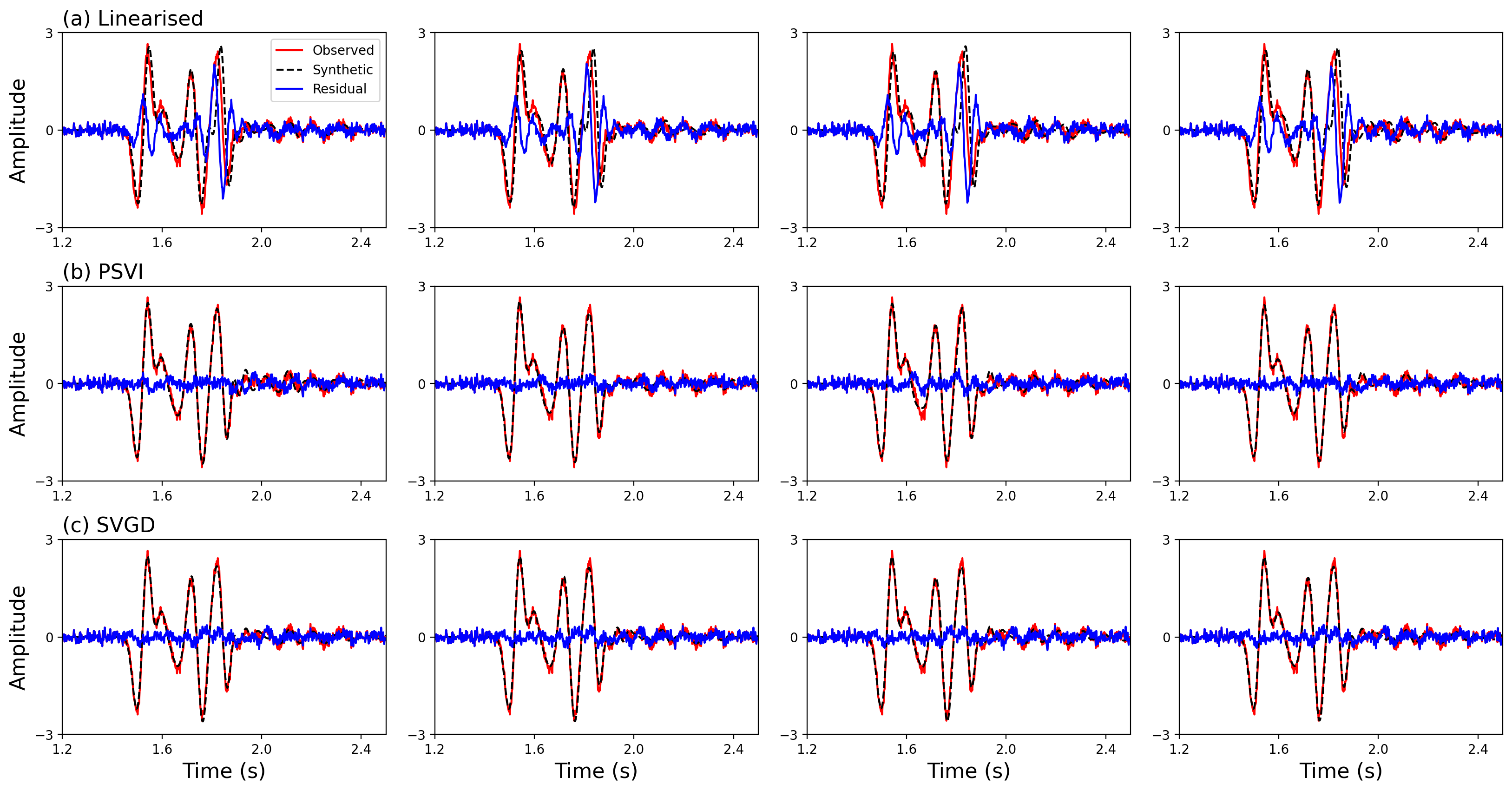}
	\caption{Comparison of one observed seismogram (red curve) and the corresponding synthetic seismograms (dashed black curves) simulated for four posterior samples from (a) linearised method, (b) PSVI and (c) SVGD. Key as in Figure \ref{fig:marmousi_data_trace625_comparison}. Note that waveform data are compared between 1.2 s and 2.5 s, within which we observe clear signals. Each panel corresponds to that in Figure \ref{fig:marmousi2_data_residual_from_4samples}.}
	\label{fig:marmousi2_data_trace625_comparison_4samples}
\end{figure}

In this Appendix we compare observed waveform data used in the Marmousi example (Figure \ref{fig:marmousi_observed_data_residual}a) with data simulated from four additional posterior model samples obtained from the linearised, PSVI and SVGD inversion results discussed in the main text. Figures \ref{fig:marmousi2_data_residual_from_4samples}a, b and c display data residuals calculated from the four posterior samples from the three methods, with each panel representing data residuals from one single posterior sample. These results correspond to those in Figures \ref{fig:marmousi_observed_data_residual}b, c and d in the main text. For a clearer comparison, in Figure \ref{fig:marmousi2_data_trace625_comparison_4samples} we compare data fit of the same seismogram (between 1.2 s and 2.5 s within which clear signals are observed) as that in Figure \ref{fig:marmousi_data_trace625_comparison}. Figures \ref{fig:marmousi2_data_residual_from_4samples} and \ref{fig:marmousi2_data_trace625_comparison_4samples} demonstrate that data fits from posterior samples from the two nonlinear methods are generally better than those from the linearised method.

\label{lastpage}
\end{document}